\documentclass{article}
\def\twocol{"n"}

\usepackage[utf8]{inputenc}
\usepackage{graphicx}
\usepackage{natbib}
\usepackage{hyperref}
\usepackage{xr}

\usepackage[english]{babel}
\usepackage{float}
\usepackage{amsmath}
\usepackage{fullpage}
\usepackage{caption}
\usepackage{ifthen}
\usepackage[format=hang]{subfig}
\usepackage{threeparttablex}
\usepackage{tabularx}
\usepackage{longtable}
\usepackage{booktabs}
\usepackage{geometry}
\usepackage{lscape}
\usepackage{afterpage}
\usepackage{xcolor}

\setcitestyle{authoryear,open={(},close={)}} 
\setcitestyle{citesep={;}}

\graphicspath{ {images/} }

\begin{document}

\title{Effect of magnetospheric conditions on the morphology of Jupiter's UV main auroral emission, as observed by Juno-UVS}
\author{
    L. A. Head,
    D. Grodent,
    B. Bonfond,
    A. Moirano,
    B. Benmahi,\\
    G. Sicorello,
    J-C G\'{e}rard,
    M. F. Vogt,
    V. Hue, 
    T. Greathouse,\\ 
    G. R. Gladstone, 
    Z. Yao
}

\date{\today}

\abstract{
    Auroral emissions are a reflection of magnetospheric processes, and, at Jupiter, it is not entirely certain how the morphology of the UV main emission (ME) varies with magnetospheric compression or the strength of the central current sheet.
    This work leverages the observations from Juno-UVS to link ME variability with particular magnetospheric states.
    Novel arc-detection techniques are used to determine new reference ovals for the ME from perijoves 1 through 54, in both hemispheres, and analyse how the size and shape of the ME vary compared to this reference oval.
    The morphology and brightness of the ME vary in local time: the dawn-side ME is typically expanded and the dusk-side ME typically contracted compared to the reference oval, and the dusk-side ME being typically twice as bright as the dawn-side ME.
    Both the northern and southern ME, and the day-side and night-side ME, expand and contract from their reference ovals synchronously, which indicates that the variable size of the ME is caused by a process occurring throughout the jovian magnetosphere.
    The poleward latitudinal shift of the auroral footprint of Ganymede correlates with the poleward motion of the ME, whereas a similar relation is not present for the footprint of Io.
    Additionally, the expansion of the ME correlates well with an increase in magnetodisc current.
    These two results suggest that a changing current-sheet magnetic field is partially responsible for the variable size of the ME. 
    Finally, magnetospheric compression is linked to a global ME contraction and brightening, though this brightening occurs predominantly in the day-side ME.
    This observation, and the observation that the dusk-side ME is typically brighter than the dawn-side ME, stands in contrast to the modelled and observed behaviour of field-aligned currents and thus weakens the theoretical link between field-aligned currents and the generation of the auroral ME. 
}

\maketitle

\section{Introduction}
The ultraviolet aurorae of Jupiter, the brightest aurorae in the Solar System, are typically divided into three parts.
The most prominent of these parts comprises the main emission (ME), an approximately continuous band of emission that circumscribes Jupiter's northern and southern magnetic poles \citep{grodent:2015}.
The polar emission comprises the emission poleward of or interior to the ME.
Likewise, the outer emission describes the auroral emission equatorward of or exterior to the ME, and is dominated by diffuse auroral emission and injection signatures \citep{dumont+:2018}.
The outer emission also contains the auroral footprints of the three closest Galilean moons (Io, Europa, and Ganymede), which arise from the relative motion of the moon compared to the magnetospheric plasma and the thus-generated Alfv\'{e}n wings which propagate along the magnetic field lines connecting the moon with Jupiter's ionosphere \citep{kivelson+:1996,saur:2004}.
Much of Jupiter's magnetospheric plasma comes from its moon, Io, which is the source of some 600-2600 kg s$^{-1}$ of neutral material \citep{bagenal+dols:2020} around half of which becomes ionised \citep{bagenal+delamere:2011}, mainly via electron collisions between slowly moving neutral atoms and the corotating plasma of the Io torus \citep{bolton+:2015}.
The ME has been observed to be variable in both size and morphology. 
Thus far, understanding of this morphological variation has been limited by a lack of images of the full aurora, particularly on the night side.
However, the Juno probe and its UltraViolet Spectrograph (UVS) instrument, in a polar orbit around Jupiter since 2016, have sent back a wealth of image data with unprecedented spatial resolution that capture the shape and brightness of the full UV aurora, in both hemispheres, on both the day and night sides simultaneously \citep[e.g.][]{bonfond+:2017,greathouse+:2021}.
The aim of this work is to leverage these image data to investigate how the morphology of the ME varies in response to magnetospheric conditions and to place this variation in the context of existing theories for the generation of the ME. 

The definition of the ME is fundamentally empirical; it is based on the visual appearance of the aurora, in which a bright and approximately continuous band of emission is commonly visible in both hemispheres, though both the brightness and continuity of this band can vary greatly over time.
However, taking the average of a set of auroral images will typically reveal a bright and distinct ME.
This empirical definition does not therefore imply a single magnetospheric origin for the entirety of the ME, but rather allows for a multitude of origins for the various substructures \citep[e.g.][]{sulaiman+:2022} that compose the ME.
This is in contrast to auroral structures such as injection signatures and the moon footprints, which are defined according to the magnetospheric processes that give rise to them.
Additionally, the similar-yet-distinct term ``main oval'' can occasionally lead to confusion.
In this article, the term ``main emission'' refers to the observed bright loop of auroral emission present in any given image. 
The term ``main oval'' typically refers to the average size and position of the ME in the ionosphere.
In this work, to avoid confusion, the term ``reference oval'' will be used to denote the average ME contour.

The ME has been thought to arise from field-aligned potentials originating from the breakdown of rigid plasma corotation at a distance of around 30 R$_J$ into Jupiter's magnetosphere \citep{hill:1979,cowley+bunce:2001}.
In this framework, a field-aligned current (FAC) loop that passes from the ionosphere through the equatorial current sheet (ECS) and back to the ionosphere spins up magnetospheric plasma to corotation with Jupiter's magnetic field via a $\textbf{J} \times \textbf{B}$ force, transferring momentum from the ionosphere to the magnetosphere in the process \citep{hill:1979}. 
Since new, outward-flowing magnetospheric plasma is constantly produced from Io's plasma torus, there must be a permanent transfer of angular momentum to ensure corotation, since angular momentum of the plasma decreases as the plasma moves outwards from Jupiter.
At the distance that the current system can no longer support the demand for angular momentum, due to limits imposed by the conductivity of the ionosphere and the rate of mass outflow from the Io torus, this rigid corotation breaks down \citep{hill:1979}, whereby field-aligned potentials originate \citep{ray+:2009} that accelerate electrons in Jupiter's ionosphere, producing the ME.
Observations support several aspects of this model; Jupiter's magnetosphere has been observed to depart from rigid corotation at 20 R$_J$ \citep{belcher+:1980} and models predict a ME that is consistent with observations in both global brightness and position \citep{cowley+bunce:2001}. 
However, while the field-aligned currents detected by the Juno spacecraft have intensities sufficient to account for the power emitted by the ME \citep{nichols+cowley:2022, kamran+:2022}, these currents are fragmented and north-south asymmetric in the magnetosphere \citep{kotsiaros+:2019}.
Additionally, the field-aligned potentials, theorised to produce a unidirectional precipitation of electrons and hence the ME in the corotation-enforcement-current model, have only rarely been detected by Juno, and, even here, the distribution of precipitating electrons remains decidedly bidirectional \citep{mauk+:2018,sulaiman+:2022} and their acceleration dominated by stochastic processes \citep{salveter+:2022}.

An alternative generation mechanism for the ME based on Alfv\'{e}n waves has also been proposed \citep{saur+:2003}.
In this model, small-scale perturbations of the ECS plasma (Alfv\'{e}nic turbulence) in the middle magnetosphere propagate along magnetic field lines in the form of Alfv\'{e}n waves. 
These waves can be partially reflected when they encounter plasma-density gradients, most notably the ionosphere boundary at one end of their path and the magnetodisc at the other.
The non-linear interaction between these rebounding waves prompts a turbulent cascade toward shorter wavelengths. 
Once the wavelength becomes comparable to the kinetic scale of magnetospheric particles, these cascading Alfv\'{e}n waves can undergo wave-particle interactions via Landau damping \citep{saur+:2018} to transfer energy to the magnetospheric plasma and hence accelerate auroral particles onto the ionosphere.
This acceleration is stochastic rather than unidirectional, which is consistent with the bidirectional electron distributions seen above the aurora by Juno \citep{mauk+:2017}. 
This framework also predicts Poynting fluxes sufficient to power the ME \citep{saur+:2018}, and the emitted power of the ME appears to correlate well with the measured intensity of Ultra-Low Frequency (ULF) waves in the jovian magnetosphere \citep{pan+:2021}.
Additionally, while the observed relationship between ME energy flux and characteristic electron energy in the ME \citep{gustin+:2004,gerard+:2016} is consistent with predictions of the corotation-enforcement-current model \citep{knight+:1973}, a similarly compatible relationship is also predicted under the Alfv\'{e}nic framework \citep{clark+:2018}. 
Alfv\'{e}nic activity at high latitudes has also been observationally confirmed, with sufficiently high Poynting fluxes to power auroral emissions \citep{lorch+:2022}.
Both the corotation-enforcement-current model \citep{cowley+bunce:2001} and the Alfv\'{e}nic model \citep{saur+:2003} predict a ME location consistent with observations.
It is currently unclear to exactly what extent these two mechanisms contribute toward the generation of the ME.


Recent observations of the ME have noted several aspects that are not predicted by a FAC-based origin. 
Firstly, models of the magnetospheric current system predict a strong day-night asymmetry in the density of FACs, with the azimuthal current density expected to be far greater at night than during the day \citep{khurana:2001,chane+:2017}, as well as lesser dawn-dusk asymmetry, in which the dawn-side ME is expected to be brighter than the dusk-side ME by an order of magnitude due to the increased bendback of the magnetic field, the increased radial current \citep{khurana:2001}, and the strengthened FACs that this would entail \citep{khurana:2001,ray+:2014}.
These asymmetries in FAC density have been confirmed observationally by Juno \citep{lorch+:2020}.
However, the dominant asymmetry in the brightness of the ME appears to be the dawn-dusk asymmetry, not a day-night asymmetry, and is inconsistent with the models and observation sof the distribution of FACs in the magnetosphere: the dusk-side ME emits around four times as much power as the dawn-side ME \citep{bonfond+:2015_1,groulard+:2024}.
This may indicate that FACs do not contribute straightforwardly to the brightness of the ME. 
Under the Alfv\'{e}nic framework, this would correspond to a greater degree of turbulence in the dusk-side middle magnetosphere, which is indeed supported by Galileo magnetometer measurements \citep{tao+:2015}.
Additionally, the modelled response of the FACs under conditions of magnetospheric compression by the solar wind does not appear to align with observations of the brightness of the ME.
For example, the models of both \citet{chane+:2017} and \citet{sarkango+:2019} indicate that the day-night asymmetry of the FAC density should increase under conditions of solar-wind compression; the model of \citet{sarkango+:2019} predicts an absolute drop in day-side FAC density and hence a supposed decrease in day-side ME brightness.
Instead, the ME appears to increase in day-side brightness during compression of the magnetosphere \citep{yao+:2022}.

The ME is considered to be a comparatively steady structure that does not rapidly change its size or brightness, unlike the polar emission, which contains structures that can appear and disappear on timescales of seconds, such as flashes \citep{palmaerts+:2023}.
However, it has been previously determined that the size and brightness of the ME are, in fact, neither static nor continuous, but instead vary in local time \citep{grodent+:2003} and in response to conditions present within Jupiter's magnetosphere \citep{bonfond+:2012,tao+:2018} and in the interplanetary medium \citep{nichols+:2017, yao+:2022}.

In addition to the previously discussed dawn-dusk asymmetry in brightness, the ME typically also shows a pre-noon discontinuity in its morphology \citep{radioti+:2008}, attributed to a persistent thermal-pressure minimum and corresponding reduced plasma-velocity gradient, caused by the interaction between the rotating plasma of the magnetosphere with the magnetopause \citep{khurana:2001,chane+:2013,chane+:2018}.
The general morphology of the ME is also known to vary with local time, with the dusk-side ME being far more disrupted and discontinuous than the dawn-side ME \citep{nichols+:2009,palmaerts+:2023}. 
Additionally, it has been suggested that the ME tends to be contracted on the dusk side and expanded on the dawn side, compared to its nominally fixed position in System-III longitude, though this conclusion was heavily biased in viewing geometry \citep{grodent+:2003}.
These results, that the ME shows brightness profiles and morphologies that depend on local time, imply that the solar wind can exert an influence deep into Jupiter's middle magnetosphere \citep{khurana:2001}, whence the ME is expected to originate. 
Hubble-Space-Telescope (HST) observations of the day-side brightness of the ME indicate a positive correlation with solar-wind pressure, as measured by Juno during its approach toward Jupiter \citep{nichols+:2017}, which aligns with the general brightening of the day-side aurora with increased solar-wind pressure observed by the Hisaki-EXCEED telescope \citep{kita+:2016}.
\citet{yao+:2022} investigated the state of compression of the magnetosphere directly, by using the detection of trapped low-frequency radio continuum radiation by Juno as a marker for magnetopause traversal, and found that global brightening of the ME systematically occurs during periods of magnetospheric compression.

There are also sources of variability in the brightness and morphology of the ME that are internal to the jovian magnetosphere, such as the variable rate of mass outflow from the Io plasma torus. 
The brightness of the northern aurora was observed to increase during a period of increased torus brightness and Io volcanic activity in 2015 \citep{tao+:2018}.
This was suggested to arise from an increase in the strength of FACs from the increased plasma mass outflow rate \citep{nichols:2011} implied by Io's greater degree of volcanism.
Increased mass outflow rate from the Io torus is also expected to lead to a higher plasma-sheet density and hence larger azimuthal currents in Jupiter's ECS, which, in turn, increases contribution of the ECS to the global magnetic field \citep{hill:2001,bonfond+:2012}. 
This works to stretch the magnetic field of Jupiter outwards, and with it the ME equatorwards.
This is in agreement with observations showing an equatorward expansion of the ME during another period of enhanced volcanic activity on Io \citep{bonfond+:2012}, though the extent to which Io's volcanism can be related to the loading of the plasma torus is disputed \citep{roth+:2020,bagenal+dols:2020}.
In general, it is not well understood to what extent these two sources (solar wind, mass outflow from the torus) contribute to the observed morphological variability of the ME.

Additionally, the variable size of the ME may be linked to two intermediate causes in the magnetosphere: a change in the magnetic-field topology (which changes the magnetic mapping between the ME source region and the ionosphere) or a change in the magnetospheric depth of the ME source region \citep{grodent+:2008,vogt+:2022}. 
In the first case, it would be expected that even auroral features with fixed magnetospheric source depths (such as the moon footprints) would show an expansion or contraction that correlates with the expansion or contraction of the ME.
In the second case, the moon footprints would not necessarily move with the ME, unless the reconfiguration of the ME source region itself altered the morphology of the magnetic field.
Thus, if the moon footprints are not observed to move with the expansion of the ME, it is likely that the ME source region is changing, whereas, if the footprints are indeed observed to move with the ME, it is likely that the magnetic-field morphology is variable, though a variable ME source region would not be fully excluded.
Indeed, these two sources of variation may themselves been related \citep{bonfond+:2012}.
Previous work that attempted to determine the extent to which these two processes affect the positions of auroral features based on HST images found that the footprint of Ganymede sometimes, but not always, moves with the expansion of the ME, though this result is hampered by the large uncertainties in the planetary-limb-fitting procedure and limited size of the dataset \citep{vogt+:2022}.
Additionally, in one HST image series taken during a period of increased volcanic activity on Io, the footprint of Ganymede (GFP) appeared to be observed interior to the ME, indicating that the ME source region had moved inside the orbit of Ganymede \citep{bonfond+:2012}; however, the viewing geometry (the GFP was close to the limb) and ME morphology at the time (very faint ME in the relevant sector) mean that this observation cannot, by itself, be considered sufficient proof of a variable ME-source-region distance, especially given the lack of other such detections. 
In all, images from Juno-UVS, which do not have the same restrictions on viewing geometry nor the considerable uncertainty in the centring of Jupiter as those made by HST-STIS, may be able to clarify further the contribution of these two effects to the variable expansion of the ME.

To process the ever-increasing amount of image data from HST and Juno-UVS, automated image-analysis techniques are required, such as arc detection.
This comprises a range of methods that aim to detect lines (curved or otherwise) in images. 
It has been widely applied to the analysis of Earth's aurorae, in the context of auroral-arc detection in all-sky-camera images \citep{syrjasuo+pulkkinen:1999}, the tracking of auroral arcs \citep{syrjasuo+donovan:2002}, or the assignment of auroral-arc structures to a set of classes \citep{wang+:2023}.
Many of these arc-detection applications are based on the technique of skeletonisation, reducing a two-dimensional structure in an image (such as the region covered by a particular auroral arc) to a thin skeleton that best represents the morphology of the arcs.
A skeletonised representation of a set of auroral arcs is often easier to analyse than the image data itself and is less sensitive to the presence of noise in the images \citep{syrjasuo+pulkkinen:1999}. 
A skeletonised representation of auroral-arc structure is typically preferred over bounding-region representation as auroral arcs typically do not have clearly defined borders and are characterised by a gradual decrease in intensity away from a central axis. 
Another advantage of skeletonisation is that it allows for consideration of auroral morphology independently of auroral brightness, at least for features with brightnesses above the background level that render them detectable by the skeletonisation algorithm.
This allows the relationships between auroral morphology and brightness to be investigated in a more objective fashion that by direct comparison of auroral images.
Dimmer auroral features, such as those found in the jovian polar aurora, can also be detected even when their low brightnesses leave them imperceptible against the much-brighter ME.

While auroral-arc detection has previously been applied to the terrestrial aurorae, this work represents the first application of auroral-arc detection to the aurorae of Jupiter. 
In this work, auroral-arc detection is applied to automatically characterise the global expansion and contraction of the ME and associate it with the conditions in Jupiter's magnetosphere, to further our understanding of the response of the morphology of the ME to conditions present in the magnetosphere.

\section{Observations}
The ultraviolet spectrograph on board Juno (Juno-UVS) operates in the 68-210 nm wavelength range and is mostly dedicated to the observations of the H$_2$ aurora on Jupiter \citep{gladstone+:2017, greathouse+:2013}. Juno is a spin-stabilized platform and spectrally resolved images are acquired by scanning the scene with the slit essentially perpendicular to the spin plane. A scan mirror at the entrance of the instrument allows it to point up to 30\textdegree\ away from that plane in both directions. The 7.2\textdegree-long slit has a dog-bone shape, being 0.2\textdegree\ wide at the borders and 0.025\textdegree\ wide at the center. Only the wide-slit data are used in this study, in order to maximise the signal-to-noise ratio of the produced spectral images. After a first radiation noise subtraction step \citep{bonfond+:2021}, every recorded photon detection event is projected onto the ellipsoid of Jupiter at an altitude of 400 km above the 1-bar level. The count-to-brightness conversion was performed according to the calibration of \citet{hue+:2019}. Hence, for each spin, the reconstructed image consists of two thin strips across the planet. These successive stripes need to be assembled in order to form an image of the UV aurora as complete and resolved as possible \citep{bonfond+:2017}. For each perijove and each hemisphere, a "master" map was created, made of the 100 successive spins as close as possible to the perijove time and which cover at least 75\% of the auroral region. Therefore, a total of 106 images are used in this work, one in the northern and southern hemisphere for the first 54 perijoves, barring perijove 2, where the spacecraft entered safe mode and no image data were collected.

HST images used in this work come from the GO-14105 and GO-14634 imaging campaigns, both using the Space Telescope Imaging Spectrograph (STIS) instrument with a strontium-fluoride filter to reduce the influence of geocoronal emissions. Images were processed into 10-second frames using the CALSTIS calibration tools from the Space Telescope Science Institute \citep{katsanis+mcgrath:1998}, converted to brightness in kilo-rayleigh (kR) assuming a colour ratio of 2.5 \citep{gustin+:2012}, and fitted to the ellipsoid of Jupiter as per \citet{bonfond+:2009}.

\section{Methods}
\label{section:methods}
The detection of arcs in a series of auroral images performed in this work proceeds in three phases:
\begin{itemize}
    \item Preprocessing: this step comprises the processing performed from the collection of the images by HST and UVS and before any detection-specific analysis is performed. Images are converted to a polar-projected format and smoothed to improve the efficiency of the arc-extraction algorithm.
    \item Extraction: this step comprises the detection of arcs in each preprocessed image. The image is template-matched with an artificial arc profile to detect those regions of the image that show ``arc-like'' shapes. The results of this template matching are used to determine the skeleton of the aurora, from where individual arcs can be extracted. 
    \item Characterisation: this step involves the extraction of a number of key properties of the detected arcs (brightness, position, ...). These properties are stored in a database, which allows auroral arcs to be more quickly analysed than directly from image data.
\end{itemize}

\subsection{Preprocessing}
\label{section:preprocessing}

For this work, images from Juno-UVS and HST were first transformed into a 1024$\times$1024-pixel Cartesian polar projection, i.e. as though viewed from above the northern geographical pole of Jupiter, with a System-III longitude of 0\textdegree\ toward the top of the image and 90\textdegree\ toward the right.
In the case of the southern aurora, the aurora is still typically displayed as though seen from the northern geographical pole ``through the planet'', as this allows for more intuitive comparison of images of the northern and southern aurora.
A 1024$\times$1024-pixel projection was chosen to ensure parity between the polar-projected images and the maps from which they are made; these maps have a resolution of 0.1\textdegree\ in both latitude and longitude, which is roughly equivalent to 100 km on the globe of Jupiter and hence consistent with the approximate 100-km-per-pixel resolution of the polar-projected image.
During this projection, it was assumed that the aurora be located at an altitude of 400 km; this projection altitude is a compromise between the moon footprints at 900 km and the ME at 250-400 km \citep{vasavada+:1999,bonfond+:2015_2}.
The advantages of this for large-scale image analysis are essentially twofold: firstly, the projection of each pixel to System-III coordinates is consistent between polar-projected images; and, secondly, the true size of features is preserved near the pole. 
Since the aurorae of Jupiter remain largely fixed in System-III coordinates \citep[e.g.][]{jupiterbook:aurorae} and are located near the geographical poles of Jupiter, the comparison of the aurora between any two images is greatly simplified.
The difference between the projection altitude and actual altitude of auroral features will introduce an error into the projected position of the aurora; however, for a typical Juno-UVS emission angle of 30\textdegree\ and an altitude difference of 200 km (i.e. projected vs assumed altitude of the ME), an error of some 115 km is introduced, which is equivalent to around one pixel in the polar-projected images used in this work.
In conjunction with the fact that Juno views the aurora from many different positions and hence that the errors are not systematic, the error introduced by the projection is considered negligible.

In the case of image data collected using STIS, collected photons are typically collated into 10-second image frames to investigate the evolution of auroral features over short timescales.
However, for this work, each STIS exposure was instead collated into a single frame, representing the pixelwise median of all 10-second frames within the exposure.
This has the disadvantage of reducing the signal from short-lived features; however, this collation has the added advantage of reducing noise in the image, important when attempting automated analysis, as is done in this work.
Features not in corotation (that move in System-III and hence also in the polar-projected images), such as the moon footprints, are not strongly filtered by this collation, provided they move sufficiently slowly.

\subsection{Extraction of auroral arcs}
\label{section:extraction}

\begin{figure}
    \centering
    \ifthenelse{\equal{\twocol}{"y"}}{
        \captionsetup[subfigure]{width=0.5\linewidth}
        \subfloat[UVS master image.]{
            \includegraphics[width=0.53\linewidth]{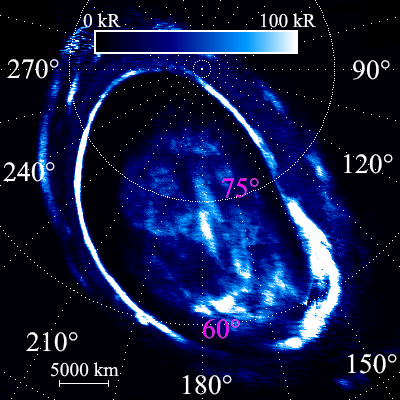}
        } \\
        \subfloat[Convolved image.]{
            \includegraphics[width=0.53\linewidth]{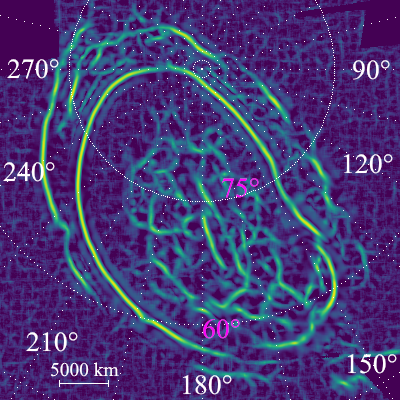}
        } \\
        \subfloat[Convolved image with noise removal.]{
            \includegraphics[width=0.53\linewidth]{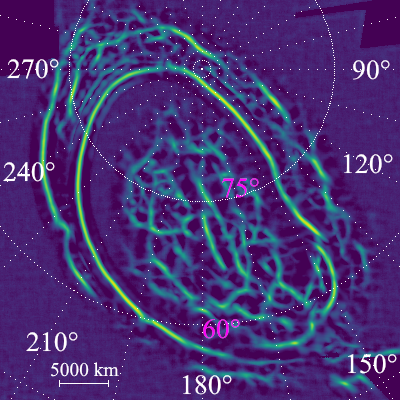}
        } \\
        \subfloat[Detected auroral arcs.]{
            \includegraphics[width=0.53\linewidth]{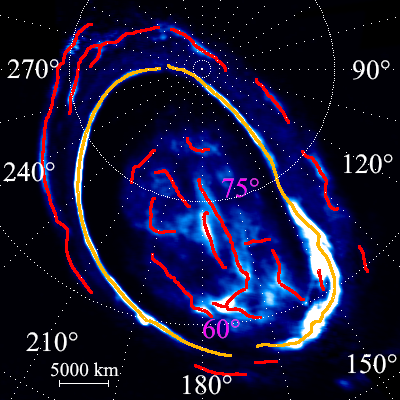}
        }
    }
    {
        \captionsetup[subfigure]{width=0.42\linewidth}
        \subfloat[UVS master image.]{
            \includegraphics[width=0.45\linewidth]{extraction_a_crop.png}
        } 
        \subfloat[Convolved image.]{
            \includegraphics[width=0.45\linewidth]{extraction_b_crop.png}
        } \\
        \subfloat[Convolved image with noise removal.]{
            \includegraphics[width=0.45\linewidth]{extraction_c_crop.png}
        } 
        \subfloat[Detected auroral arcs. ME arcs in yellow.]{
            \includegraphics[width=0.45\linewidth]{extraction_d_crop.png}
        }
    }
    \caption{
        The arc-detection algorithm applied to the UVS master image of the northern aurora from perijove 6. A 15\textdegree-by-15\textdegree\ grid in System-III longitude and planetocentric latitude is overlain on the aurora. The System-III longitude of certain gridlines are given in white, and the planetocentric latitudes of certain gridlines in magenta. The brightness scale of the images of the aurora in kR is given at the top of (a).
    }
    \label{fig:extraction}
\end{figure}

The aurorae of Jupiter are composed of multiple smaller discrete features \citep{grodent:2015}, many of which show arc-like morphologies. 
Previous work \citep[e.g.][]{vogt+:2022} tends to employ techniques that use the position of the peak in the approximately Gaussian profile of the ME around some central location in the aurora to determine its position in the images.
A natural extension of this technique would be an arc-detection algorithm that works for all arc-like structures in the aurora, even undetected arcs for which a suitable ``central point'' has not been determined.
The goal of this arc-detection algorithm, therefore, is the automatic extraction and characterisation of these auroral arcs, without bias toward brighter arcs, such as those found in the ME.
To this end, template matching with an artificial auroral-arc profile can be used to provide a measure of ``arcness'' for each pixel in the polar-projected images. 
The template element was a 13$\times$13-pixel kernel, with a normalised Gaussian profile with a FWHM of 8 px centred along the vertical axis of the kernel. 
By rotating the image between 0\textdegree\ and 180\textdegree\ in 1\textdegree\ increments, performing the template matching against the vertical arc-profile kernel using the \texttt{match\_template} function in Python3's \texttt{scikit-image} library \citep{scikit-image} on each rotated image, and taking the maximum normalised response to the template matching at each pixel over all the rotations, an arcness map of the original auroral image can be obtained that is not dependent on the brightness of the auroral arcs, as in Fig. \ref{fig:extraction}b.
However, background shot noise in the auroral images leads to the presence of an unwanted background pattern in the convolved arcness maps. 
Since the template matching produces normalised responses, the intensity of this background pattern is comparable to the intensity of the ``true'' auroral arcs, despite the considerably greater brightness of the latter in the auroral images, and would interfere with the results of later arc-detection procedures.
This background pattern is dependent on the distribution of the shot noise, and, as such, neither a change in the parameters of convolution element nor repeating the template matching will diminish its presence.
To remove this background pattern, the background noise level of the polar-projected auroral images was estimated by taking either the 50th-percentile (HST) or 75th-percentile (UVS) pixel value of those pixels with values greater than 0; the choice of percentile is a consequence of the different typical coverages of HST and UVS polar-projected images.
By inserting random Gaussian noise with a mean of 0 and a standard deviation equivalent to this background level, the distribution of the background shot noise can be modified without disturbing the much-brighter auroral arcs.
The template matching was performed 30 times with different random background noise profiles, and the final arcness map of an auroral image is the pixelwise median of these 30 iterations; see Fig. \ref{fig:extraction}c, where the background pattern is clearly diminished compared to Fig. \ref{fig:extraction}b.
Some noise can still be seen in brighter parts of the aurora. 
For this work, which concentrates on the clear arc-like profiles of the ME with a considerable manual element, this noise does not materially impact the results; however, if dimmer arcs, such as the polar filaments \citep{nichols+:2009}, were to be investigated using this algorithm, the background-noise removal process would be refined to remove this pattern from the relevant regions of the aurora. 

In the case of the aurora, it is more sensible to extract the central axes of the auroral arcs via skeletonisation, rather than attempt to define the bounding shape, as auroral arcs are not structures with well-defined borders.
To this end, a suitable threshold was applied to the arcness maps; it was determined a posteriori that estimating the background level $BG$ using the same pixel-value percentile as before, this time applied to the arcness maps, then calculating $BG + 0.25(1-BG)$ provides a suitable threshold to distinguish between true arcs and background noise.
Applying this threshold to the arcness map returns a mask of the approximate arc locations in the aurora, which can be further reduced to a skeleton of the aurora through the use of the \texttt{skeletonize} function included in the scikit-image Python library \citep{scikit-image}.
This auroral skeleton can be interpreted in the form of a mathematical graph, with nodes and edges, using the \texttt{sknw} Python library under the ImagePy framework \citep{wang+:2018}, which allows for easier modification and processing of the detected auroral arcs.
A number of processing steps are applied to this skeleton graph to extract individual instances of auroral arc:
\begin{itemize}
\item Very short arcs (fewer than 10 pixels) are removed, as these are likely to be background noise.
\item Graph nodes between three edges are assumed to be the intersection between one larger arc and one smaller arc, as it is unlikely that three independent auroral arcs would meet in the same location in the aurora. To this end, the two edges which best align at the node are associated to the same auroral arc and are hence merged together.
\item Edges that exhibit very small local radii of curvature (< 5 px) are assumed to consist of two auroral arcs that have been mistakenly included in the same edge, since auroral arcs are presumed to curve relatively gradually. As such, these edges are split into two arcs at the point of minimum radius of curvature.
\item If two arcs are well aligned and their ends close to one another, they are assumed to form part of one larger arc than has been mistakenly split into two edges and are hence merged together.
\end{itemize}
At the end of this processing, the set of detected individual auroral-arc instances is returned; see Fig. \ref{fig:extraction}d.


\subsection{Characterisation of auroral arcs}
To perform further analysis on arc-like structures in Jupiter's aurorae, it is necessary to extract key properties of the detected arcs. 
Many of these properties (such as arc brightness, position, ...) can be trivially retrieved by reprojecting the detected arcs onto the polar-projected auroral images.
However, several arc properties have slightly more involved derivations.

In much the same way that a representative ``master'' auroral brightness map can be constructed for each perijove, as in section \ref{section:preprocessing}, a master auroral colour-ratio map can be established from the ratio between the radiance at 155-162 nm and at 125-130 nm \citep{bonfond+:2017}.
Whereas the auroral brightness contains information about the flux of the precipitating electrons, the colour ratio can be used as a proxy for the electron energy; higher-energy electrons are expected to be able to penetrate further into Jupiter's atmosphere, where flux in the 125-130 nm band is more strongly attenuated by the CH$_4$ layer, thus leading to a higher (``redder'') colour ratio.

\label{section:mag_field_model}
In this work, the JRM33 internal-magnetic-field model of Jupiter \citep{connerney+:2022} is used together with the Con2020 model of the external (ECS) magnetic field \citep{connerney+:2020} to model the total magnetic field, and to provide mappings between ionospheric and presumed ECS locations of detected auroral arcs. 
An 18th-order JRM33 magnetic-field fit is used to ensure the best-possible correspondence between the modelled and observed positions of the moon footprints, particularly that of Ganymede \citep{moirano+:2024}.
These models are contained within the JupiterMag Python wrapper \citep{jupitermag} as part of the Magnetospheres of the Outer Planets Community Code project \citep{wilson+:2023}. 

\begin{figure}
    \centering
    \ifthenelse{\equal{\twocol}{"y"}}{
        \captionsetup[subfigure]{width=0.95\linewidth}
        \subfloat[North.]{
            \includegraphics[width=\linewidth]{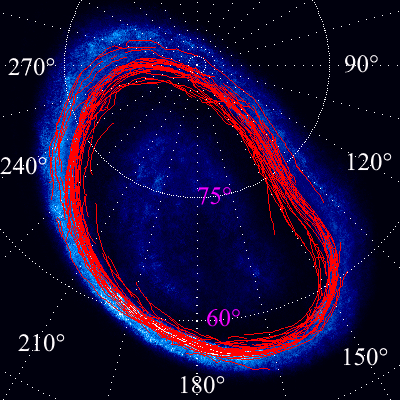}
        } \\
        \subfloat[South.]{
            \includegraphics[width=\linewidth]{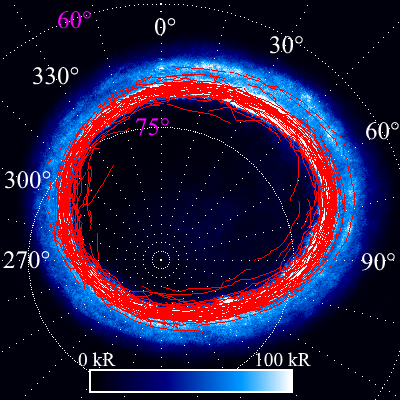}
        }
    }
    {
        \captionsetup[subfigure]{width=0.42\linewidth}
        \subfloat[North.]{
            \includegraphics[width=0.45\linewidth]{MainEmissionAnalysis_UVS_polarviewAvrOverlay_N_crop.png}
        } 
        \subfloat[South.]{
            \includegraphics[width=0.45\linewidth]{MainEmissionAnalysis_UVS_polarviewAvrOverlay_S_crop.png}
        } 
    }
    \caption{
        The detected ME arcs for perijoves 1 through 54, shown in red, overlain on the pixelwise median-average aurora for each hemisphere. A 15\textdegree-by-15\textdegree\ grid in System-III longitude and planetocentric latitude is overlain on the aurora. The System-III longitude of certain gridlines are given in white, and the planetocentric latitudes of certain gridlines in magenta. The brightness scale of the images of the aurora in kR is given at the bottom of (b).
    }
    \label{fig:detected_ME}
\end{figure}

Of particular importance to this work is the correct identification of those auroral arcs that comprise the main auroral emission. 
The ME is broadly associated with the innermost semi-continuous bright arc in the aurora, though, as discussed above, this definition is empirical and does not prescribe a single shared magnetospheric origin to the entire ME.
This morphology can be strongly disrupted by the presence of bridges \citep{palmaerts+:2023}, strong injection signatures \citep{grodent+:2018}, or dawn storms \citep{bonfond+:2021}, among other causes.
The ME can therefore have starkly different sizes and morphologies between any two images of the aurora, and, as such, designation of the approximate region of the polar-projected images that contain the ME was performed manually. 
This manual designation was based on the convolved images (such as Fig. \ref{fig:extraction}c) with reference to the original brightness images (Fig. \ref{fig:extraction}a) to ensure that only bright regions showing strong arc-like profiles (or dimmer continuations of the same bright arc) were attributed to the ME.
For each detected auroral arc, the part of the skeleton that fell within this manually designated region was taken to belong to the ME.
This method has the benefit of the unbiased detection of auroral arcs from the automatic arc-detection algorithm as well as the certainty that only the ME is considered for further analysis. 
An automatic ME designation containing those arcs within a certain distance of the average UVS main oval was a posteriori determined to be inadequate; a distance limit from the reference contour sufficiently large to capture the full range of contraction and expansion of the ME would also result in the inclusion of many detected poleward and equatorward arcs which do not form part of the ME.
Manual designation also allows for the exclusion of the most heavily disrupted regions of the ME, where the arc-detection algorithm performs most poorly. 

The results of this semi-automatic ME-detection process are given in Fig. \ref{fig:detected_ME}.
The majority of the detected ME arcs are located on or near the reference contour for perijoves 1 to 54 (see section \ref{section:uvs_ref_oval}).
The ME is also well sampled in System-III longitude; there are no portions of the ME that are distinctly under- or over-sampled.
This is as expected, as many features that would prevent proper detection of portions of the ME (the presence of bridges, dusk-side disruption, dawn storms) are not fixed in System-III longitude. 
Additionally, the average aurorae shown in Fig. \ref{fig:detected_ME} would, at first glance, not appear to demonstrate the empirical result that approximately one third of the total auroral power can be attributed to each of the three regions of the aurora (ME, polar emission, outer emission) \citep{grodent+:2018}.
This is due to the median averaging performed to obtain the average aurorae; for example, bright but transient features that dominate the power output of the polar region are disproportionately diminished by median averaging.
Median averaging is used here to ensure that the ME, the main subject of this work, is as clear as possible, free from the unwanted effects of transient elements such as dawn storms.

\section{Results and discussion}

\subsection{New reference oval for the main emission}
\label{section:uvs_ref_oval}

\begin{figure}
    \centering
    \ifthenelse{\equal{\twocol}{"y"}}{
        \captionsetup[subfigure]{width=0.95\linewidth}
        \subfloat[North.]{
            \includegraphics[width=\linewidth]{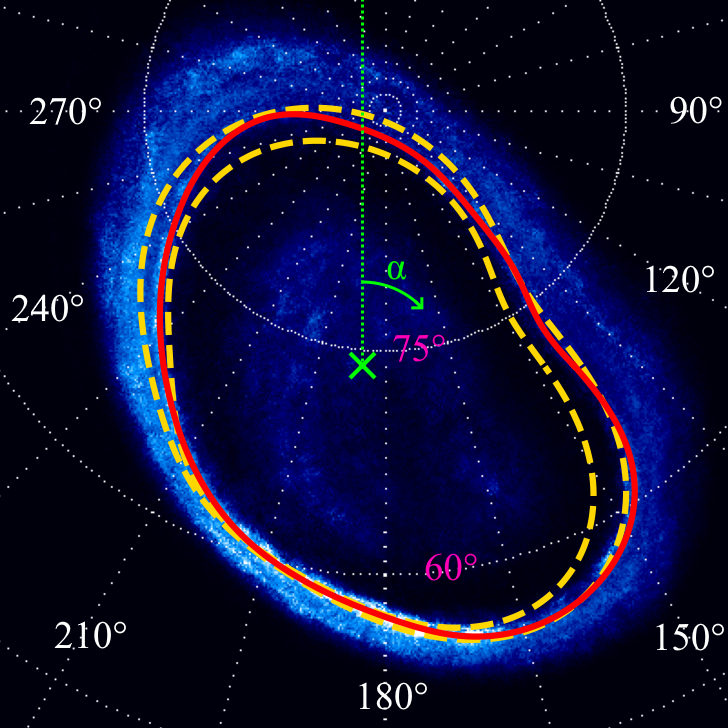}
        } \\
        \subfloat[South.]{
            \includegraphics[width=\linewidth]{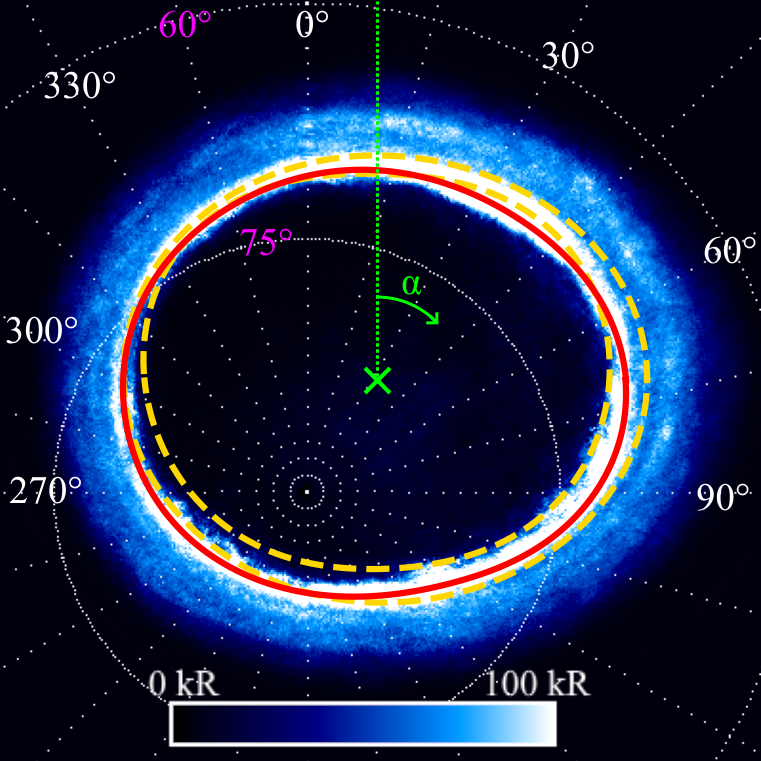}
        }
    }
    {
        \captionsetup[subfigure]{width=0.42\linewidth}
        \subfloat[North.]{
            \includegraphics[width=0.45\linewidth]{refOval_n_crop.png}
        } 
        \subfloat[South.]{
            \includegraphics[width=0.45\linewidth]{refOval_s_crop.png}
        } 
    }
    \caption{
        The ME reference ovals defined in this work, overlain on the pixelwise median-average aurora for each hemisphere. The new UVS reference oval is shown as a solid red line. The expanded and contracted HST ME reference ovals from \citet{bonfond+:2012}, shown as dashed yellow lines, are included for comparison. The pseudo-magnetic-coordinate reference point is denoted by a green cross, alongside the sense of pseudo-magnetic angle $\alpha$. A 15\textdegree-by-15\textdegree\ grid in System-III longitude and planetocentric latitude is overlain on the aurora. The System-III longitude of certain gridlines are given in white, and the planetocentric latitudes of certain gridlines in magenta. The brightness scale of the images of the aurora in kR is given at the bottom of (b).
    }
    \label{fig:ref_oval}
\end{figure}

Any investigation of the variable size of the ME must necessarily define a reference ME profile against which individual images of the ME can be compared. 
There exist previously defined ME reference ovals in the literature \citep[e.g.][]{bonfond+:2012}; however, for this work, new reference ovals were defined based on the average position of the ME in the UVS master images between perijoves 1 and 54.
These master images were stacked and the pixelwise median brightness (for those pixels with UVS coverage) taken to produce an average ME profile, separately for both the northern and southern hemispheres; see Fig. \ref{fig:ref_oval}.
This average ME profile was then convolved with a Gaussian kernel, as described in section \ref{section:extraction}, to provide a smooth and continuous contour for the average ME.
A central reference point was defined for each hemisphere (left-handed System-III longitude $\phi_{S3}=185$\textdegree, planetocentric latitude $\theta=74$\textdegree\ in the north; $\phi_{S3}=32$\textdegree, $\theta=-82$\textdegree\ in the south; taken from \citet{bonfond+:2012}) and used to define a pseudo-magnetic coordinate (radius and angle around the reference point) for each point in the ME profile.
A univariate spline was fitted through the pseudo-magnetic coordinates of the ME, then converted to left-handed System-III longitude and planetocentric latitude to provide the final ME reference oval for each hemisphere.
The use of pseudo-magnetic coordinates is preferred here over System-III longitude and planetocentric latitude to ensure a sensible spline fitting that evenly samples the contour of the ME, most necessary in the northern hemisphere where the ME deviates noticeably from its idealised circular shape.

It can be seen in Fig. \ref{fig:ref_oval} that the new reference ovals well describe the average position of the ME between perijoves 1 and 54 in both the northern and southern hemispheres. 
For the most part, they also fall within the range of ME positions determined from HST data by \citet{bonfond+:2012}.
The UVS reference ovals differ from the \citet{bonfond+:2012} reference ovals most notably at high latitudes; this is to be expected, as the \citet{bonfond+:2012} reference ovals are based on HST observations of the ME, in which the high-latitude aurora is rendered unobservable by the typical viewing geometry.
The \citet{bonfond+:2012} ME reference ovals also notably underestimate the size of the average UVS ME in the northern hemisphere in the region of the magnetic anomaly along the 150\textdegree\ System-III meridian \citep{grodent+:2008}.
This typically corresponds to the dusk-side hemisphere in HST images, in which the ME tends to be contracted \citep{grodent+:2003}.
It is therefore unsurprising that the HST-based \citet{bonfond+:2012} reference ovals underestimate the size of the ME in this sector. 
In the southern hemisphere, the ME in the 15\textdegree-to-75\textdegree\ longitude range typically corresponds to the dawn-side hemisphere in HST images, and it is thus equally unsurprising that the \citet{bonfond+:2012} reference ovals overestimate the size of the ME in this sector. 
These reference ovals are given in the Supplementary Material associated with this paper.

\subsection{Global behaviour of the main emission}
\begin{figure}
    \centering
    \ifthenelse{\equal{\twocol}{"y"}}{
        \includegraphics[width=\linewidth]{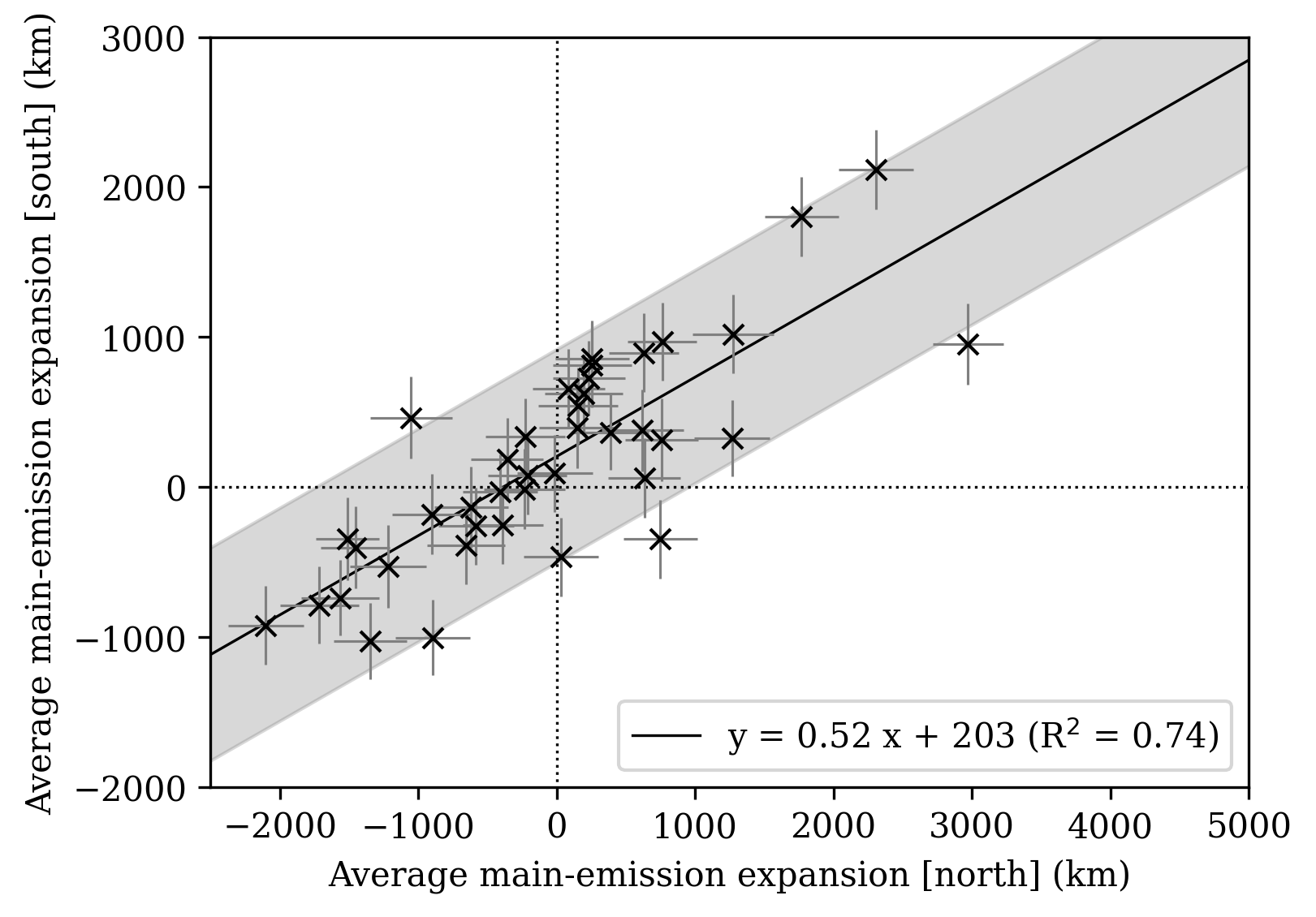}
    }
    {
        \includegraphics[width=\linewidth]{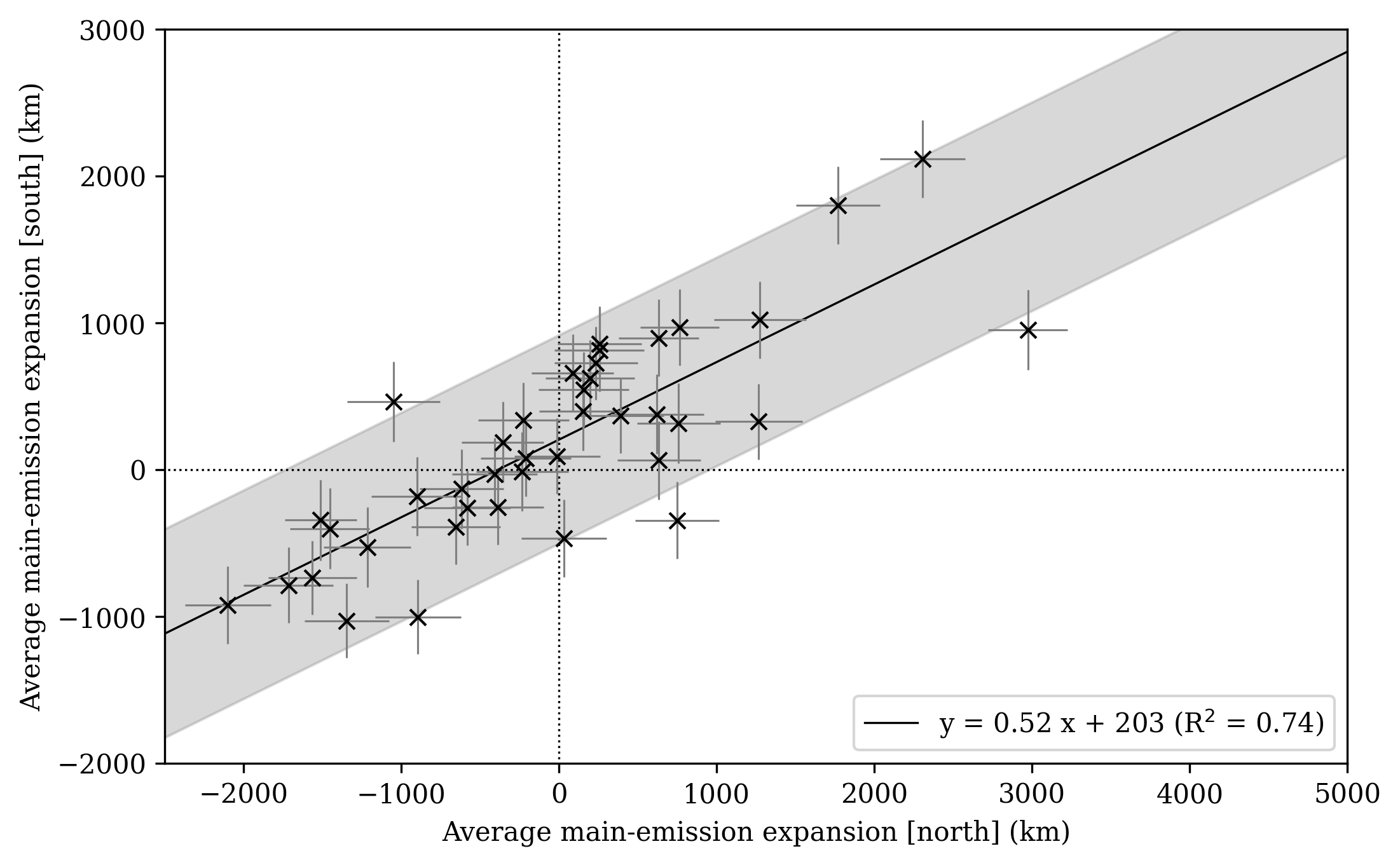}
    }
    \caption{
        The median-averaged global ME expansion from the UVS reference oval in the north vs in the south for perijoves 1 through 54. Perijoves with sufficiently poor coverage in the northern hemisphere such that the position of no part of the ME can be reliably determined are omitted. The uncertainty in the global ME expansion has been estimated from the HWHM of the Gaussian profile arc-detector convolution element (3 px), which correlates with the width of the ME arcs in the convolved image. The fitted relationship is given by a solid black line; its form and R-squared goodness-of-fit value are given in the legend. The 1$\sigma$ confidence level of the fit is given by the shaded region. The position of the origin is denoted by two dotted black lines. Negative (positive) values of ME expansion indicate a global contraction (expansion) of the ME.
    }
    \label{fig:north_south_comparison}
\end{figure}

Using the UVS reference oval and the automatically detected ME arcs as described in section \ref{section:methods}, it is possible to calculate the average global expansion of the ME for each perijove hemisphere. 
Using the same central reference point as shown in Fig. \ref{fig:ref_oval}, the detected ME arcs for each perijove hemisphere were converted to pseudo-magnetic coordinates as per section \ref{section:uvs_ref_oval}.
For each pixel in the ME arcs, the point on the reference oval with the same pseudo-magnetic angle was used to calculate the pixelwise expansion of the ME, since it is assumed that, for sufficiently small shifts, the aurora expands perpendicularly from the reference oval under conditions of global expansion or contraction.
This expansion was calculated in kilometres by projecting the ME pixel and the reference oval to the globe of Jupiter.
By taking the median pixelwise expansion of every pixel in each ME arc, an average ME expansion can be defined for each perijove hemisphere. 
This expansion is positive for an equatorward expansion of the ME and negative for a poleward contraction.
To test the assumption that the ME expand perpendicularly from its reference contour, the UVS ME reference contour was expanded and contracted by 20 pixels (approximately equivalent to 2000 km, comparable to the range of ME expansion seen in Fig. \ref{fig:north_south_comparison}).
Each point on the original ME reference contour was linked to a point on the expanded/contracted reference contour by taking the pixel with the nearest pseudo-magnetic angle. 
The expansion/contraction of each point was calculated.
The same was performed taking the pixel with the closest mapped magnetospheric System-III longitude and the expansion/contraction once again calculated.
The median difference between these two estimations of expansion/contraction for each point on the original reference contour was less than 200 km in both the northern and southern hemispheres, which is less than the estimated uncertainty of the expansion of the ME, and so the use of pseudo-magnetic coordinates to calculate the expansion of the ME is reasonable.

As shown in Fig. \ref{fig:north_south_comparison}, there is a clear (R$^2$ = 74\%) positive correlation between the global expansion of the ME in the north and the south during a given perijove. 
This implies that the physical origin of the variability in ME expansion is a global phenomenon that affects the aurorae in both hemispheres in the same manner, and indicates therefore that the process(es) giving rise to the ME are magnetospheric, and not ionospheric, in nature.
The processes controlling the expansion of the ME must therefore also vary over timescales no shorter than the time required for Juno to pass from the northern to the southern hemisphere during a perijove ($\sim$2.5 hours).
The fact that the relation passes close to the origin indicates that the average position of the ME in the two hemispheres corresponds to the same magnetospheric state. 
Fig. \ref{fig:north_south_comparison} also shows that the expansion of the northern ME varies around twice as much as the southern ME in absolute terms.
This is as expected given the presence of the low-field-strength magnetic anomaly in the north \citep{bonfond+:2015_1} and thus the same relative expansion of the ME would typically result in a larger absolute-distance expansion in the north than in the south.
There is also a similar region of elevated field strength in the northern hemisphere; however, an argument based on the conservation of magnetic flux indicates that low-field anomalies will more greatly affect the movement of the ME than high-field anomalies.
The reader is asked to imagine an event in the magnetosphere that stretches the field lines outward.
A series of field lines will pass through an arbitrary section of the ME source region; these field lines can be thought of as a flux tube, and hence the total tube magnetic flux is conserved between the ECS and the ionosphere (IS), as
\begin{equation}
    \Phi=B_{\texttt{ECS}}S_{\texttt{ECS}}=B_{\texttt{IS}}S_{\texttt{IS}},
\end{equation}
where $\Phi$ is the total magnetic flux in the flux tube, $B_{ECS}$ and $B_{IS}$ refer to the magnetic field strength, and $S_{ECS}$ and $S_{IS}$ to the flux-tube foot surface area in the ECS and ionosphere respectively.
We are free to choose rectangular flux-tube ends, such that
\begin{equation}
    S_{\texttt{ECS}}=\Delta r_{\texttt{ECS}} l_{\texttt{ECS}} , \  S_{\texttt{IS}}=\Delta r_{\texttt{IS}} l_{\texttt{IS}},
\end{equation}
where $\Delta r$ refers to the radial shift of the magnetic field lines connecting the ME in the ionosphere to the ECS and $l$ refers to an arbitrary (small) flux-tube end width.
Since we are introducing an arbitrarily small radial stretch in the magnetic field lines in the ECS (such that $B_{\texttt{ECS}}$ remains essentially unchanged and $r_{\texttt{ECS}}$ is a small constant that we define),
\begin{equation}
    B_{\texttt{IS}}S_{\texttt{IS}}=\texttt{Const.}
\end{equation}
and hence
\begin{equation}
    \Delta r_{\texttt{IS}} \propto B_{\texttt{IS}}^{-1}.
\end{equation}
Given this relation, and that the northern-hemisphere weak-field and strong-field anomalies are approximately half and twice as strong as the average surface magnetic field \citep{moirano+:2024}, it is therefore reasonable to assume that the weak-field anomaly affects the expansion of the northern ME twice as much as the strong-field anomaly in absolute distance, and hence that the northern ME would show a greater expansion in ionospheric distance than the southern ME for a given state of the magnetosphere. 

\begin{figure}
    \centering
    \ifthenelse{\equal{\twocol}{"y"}}{
        \includegraphics[width=\linewidth]{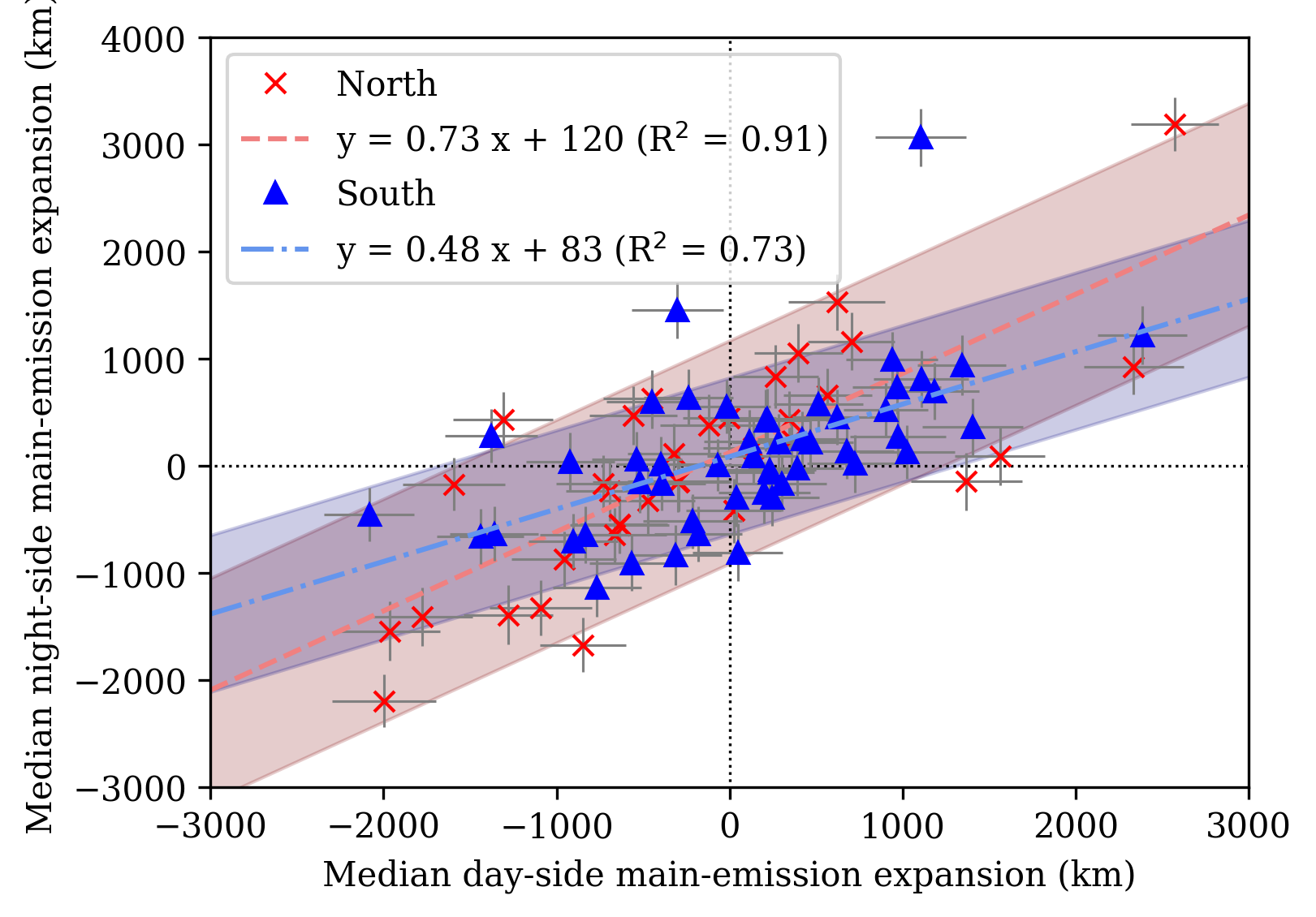}
    }
    {
        \includegraphics[width=\linewidth]{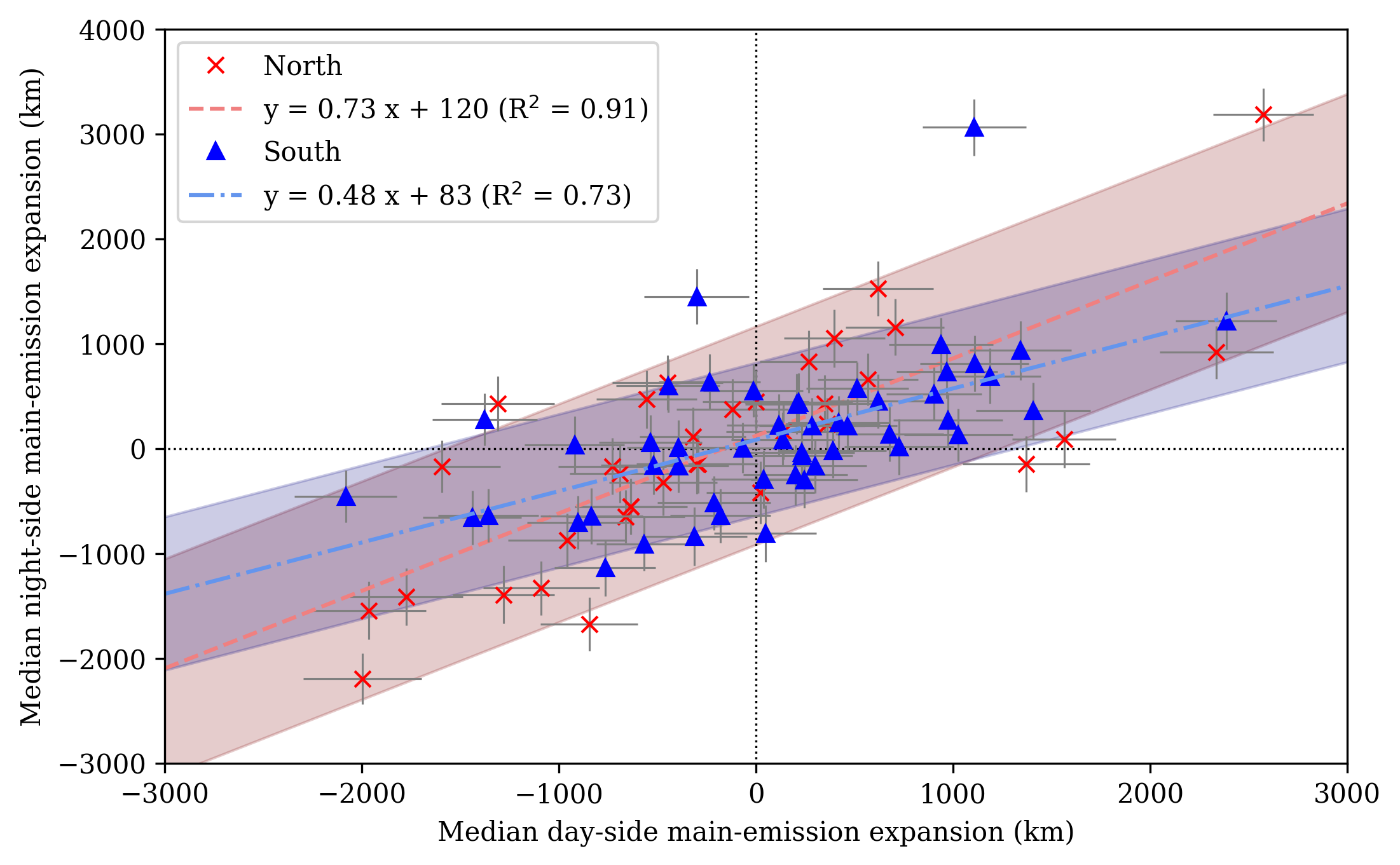}
    }
    \caption{
        The median-averaged global ME expansion from the UVS reference oval in the day-side hemisphere vs in the night-side hemisphere for perijoves 1 through 54, for the northern (red crosses) and southern (blue triangles) aurorae. Perijoves with excessively poor coverage in the northern hemisphere are omitted. The uncertainty in the global ME expansion has been estimated from the HWHM of the Gaussian profile arc-detector convolution element (3 px), which correlates with the width of the ME arcs in the convolved image. The fitted relationships are given by a red dashed line and a blue dot-dashed line for the northern and southern hemisphere respectively; their forms and R-squared goodness-of-fit values are given in the legend. The 1$\sigma$ confidence levels of the fits are given by the red and blue shaded regions, for the northern and southern hemispheres respectively. The position of the origin is denoted by two dotted black lines. Negative values of ME expansion indicate a global contraction of the ME.
    }
    \label{fig:night_day_comparison}
\end{figure}

The expansion of the ME from its reference oval can also be investigated separately for regions of the ME that magnetically map to the day-side (06:00 to 18:00 local time) and night-side (18:00 to 06:00 local time) magnetosphere.
Fig. \ref{fig:night_day_comparison} shows that, in both the northern and southern hemispheres, the correspondence between the expansion of the night-side ME and the expansion of the day-side ME can be well described by a positive linear relation (91\% and 73\% of the total variance in the data can be described by a linear relation, for the northern and southern hemisphere respectively).
When the day-side ME is contracted, so too is the night-side ME.
In both hemispheres, the day-side ME expands more than the night-side ME; in the southern hemisphere, the day-side ME expands and contracts more than twice as much as the night-side ME in absolute distance.
The fitted relations pass through the (0,0) origin in both hemispheres to within the error of the data (approximately $\pm$400 km).
In all, these results indicate that processes that work to contract the ME affect both hemispheres simultaneously, though the day-side ME is typically affected to a greater extent.

\begin{figure}
    \centering
    \ifthenelse{\equal{\twocol}{"y"}}{
        \includegraphics[width=\linewidth]{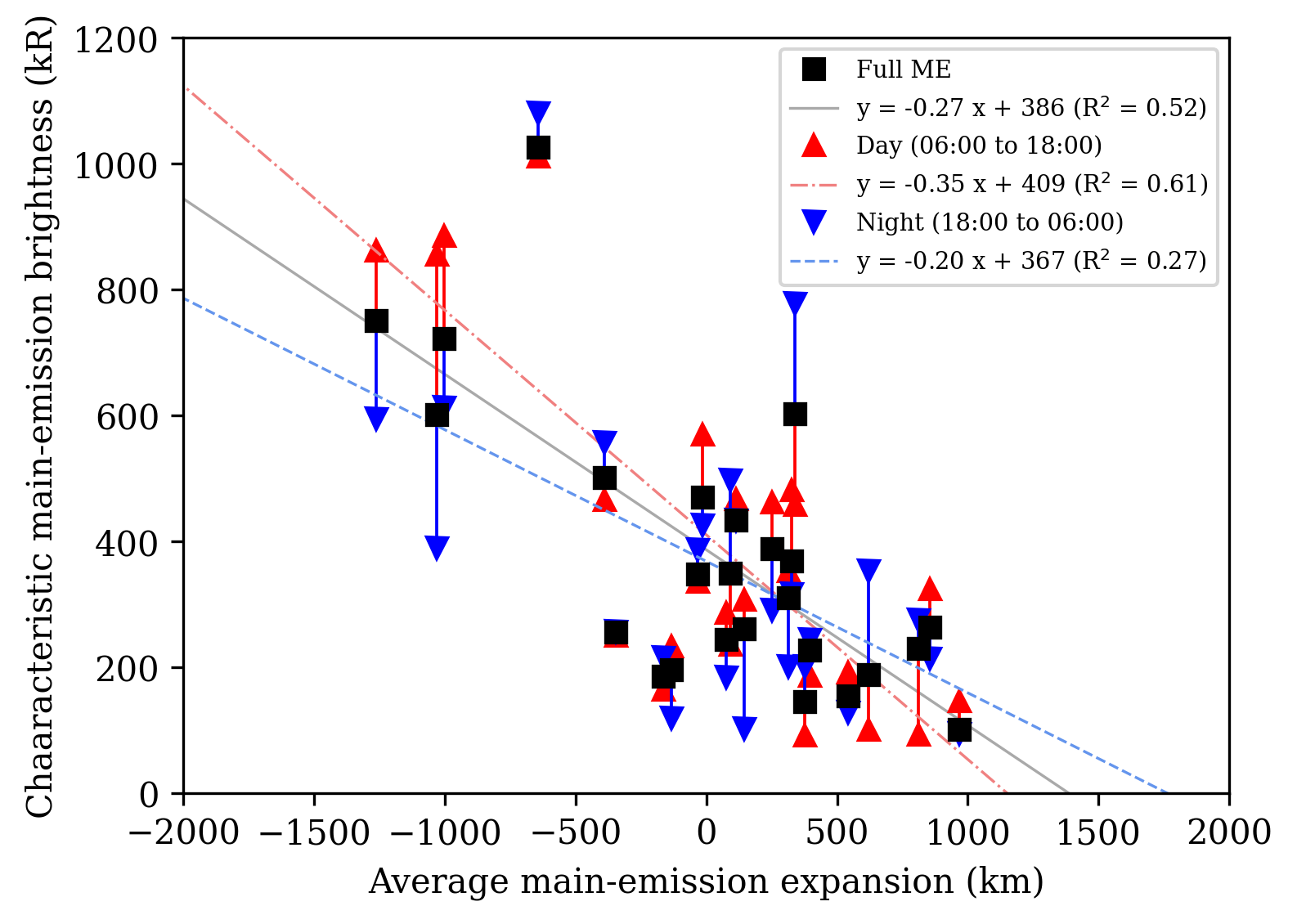}
    }
    {
        \includegraphics[width=\linewidth]{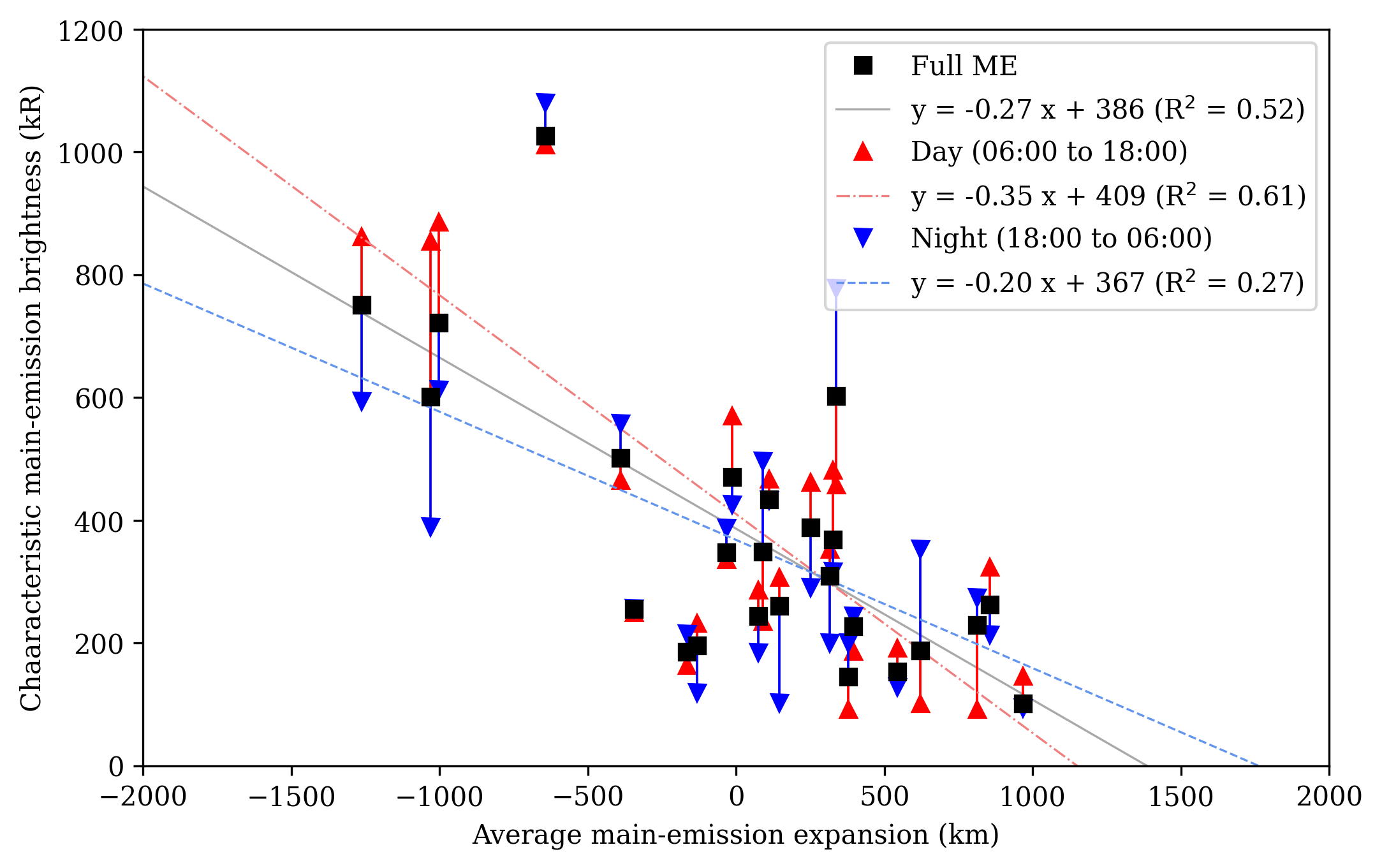}
    }
    \caption{
        The characteristic ME brightness vs the median-averaged global ME expansion from perijoves 1 through 54, in the southern hemisphere. The characteristic brightness of the entire ME is given by black squares, and the characteristic brightness of only the day- and night-side hemispheres by red and blue triangles respectively. Only those cases with a pseudo-magnetic angular ME coverage greater than 80\% have been included. Linear fits to the data have been calculated and shown as a black solid line (full ME), a red dashed line (day), or a blue dotted line (night). The form and R-squared goodness-of-fit value of these fits are given in the legend.
    }
    \label{fig:comp_vs_power}
\end{figure}

The use of the automatic arc-detection algorithm in this work permits the systematic comparison of the global ME expansion with other parameters, such as the brightness of the ME.
In this work, the median-average brightness of all pixels in the automatically detected arcs of the ME was taken as a characteristic measurement of ME brightness.
Note that the median brightness of the detected ME cannot be directly related to any ``average'' emitted auroral power measured by Juno-UVS and depends on the pixel binning of the polar projections; however, it does allow for the straightforward comparison of a characteristic brightness of the ME between images, or between different regions of the ME in the same image.
To ensure a sensible comparison, only those perijoves with a pseudo-magnetic ME coverage greater than 80\% were taken, which amounts to 8 cases in the north and 25 cases in the south.
Note that ``coverage'' refers to the total pseudo-magnetic angular coverage of detected arcs that can be reliably associated with the ME; it is possible that the entire aurora be imaged by UVS yet the coverage be less than 100\% if, for example, parts of the ME are too dim or morphologically disrupted to be unquestionably identified as belonging to the ME.
Fig. \ref{fig:comp_vs_power} shows that, in the southern hemisphere, the characteristic ME brightness increases with contraction (negative expansion) of the ME.    
The R$^2$ value of this relationship indicates that a linear response in ME brightness to ME contraction can account for 52\% of the variance in the data; given the frequent presence of additional features on the ME that can contribute significantly to the detected brightness (dawn storms, disrupted morphologies), it is noteworthy that more than half of this brightness variability can be attributed to a simple response to ME contraction.
Therefore, in an ideal ME, one without additional features superimposed, it would be expected that the contraction-brightness relationship be even clearer.
The brightness of the ME is typically within the the expected range of 50 to 500 kR as measured by HST \citep{grodent+:2003}.
The results for the northern hemisphere have been omitted from Fig. \ref{fig:comp_vs_power} for the sake of clarity and the comparatively few perijoves (8) with the necessary 80\% coverage of the ME by UVS. 

Additionally, Fig. \ref{fig:comp_vs_power} displays separately the characteristic brightnesses of the day-side and night-side ME, the regions of the ME that magnetically map to the ECS from 06:00 to 18:00 and from 18:00 to 06:00, respectively. 
This shows that the positive dependence of the brightness of the ME on the contraction (negative expansion) of the ME is far more striking in the day-side hemisphere, with a given ME contraction increasing the  day-side brightness by almost twice as much as the night-side brightness.
If the global contraction of the ME can be associated with an increased solar-wind ram pressure, as indicated by models \citep{promfu+:2022}, this result stands in opposition to the expected behaviour of a FAC-driven ME; both magnetohydrodynamical simulations \citep{chane+:2017,sarkango+:2019} and observations \citep{lorch+:2020} agree that solar-wind compression more greatly increases the density of FACs on the night-side hemisphere.
This is discussed in greater detail and in the context of further results below in section \ref{sec:interpreting_in_context}.

\subsection{Local-time dependence of the morphology of the main emission}

\begin{figure}
    \centering
    \ifthenelse{\equal{\twocol}{"y"}}{
        \includegraphics[width=\linewidth]{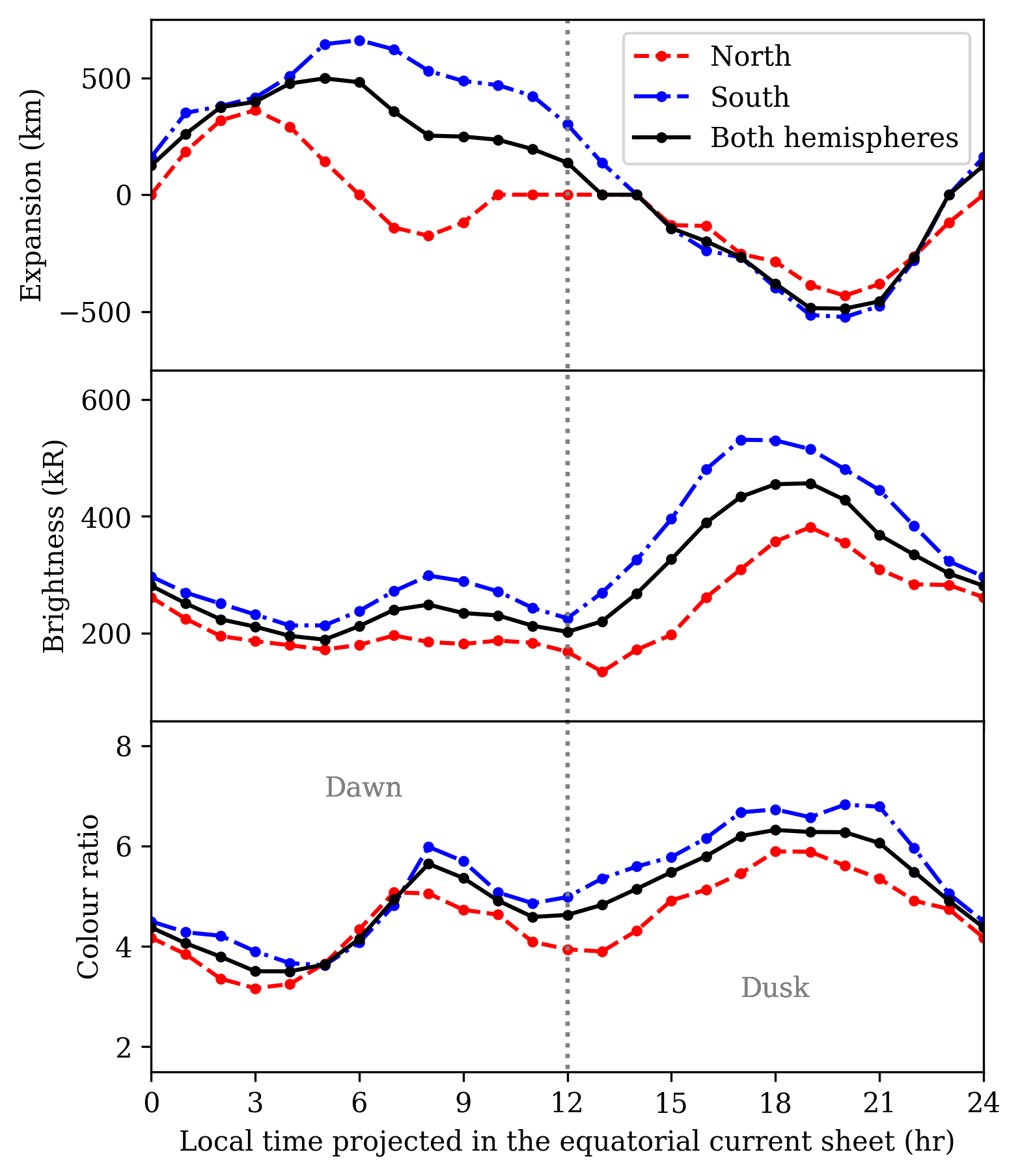}
    }
    {
        \includegraphics[width=\linewidth]{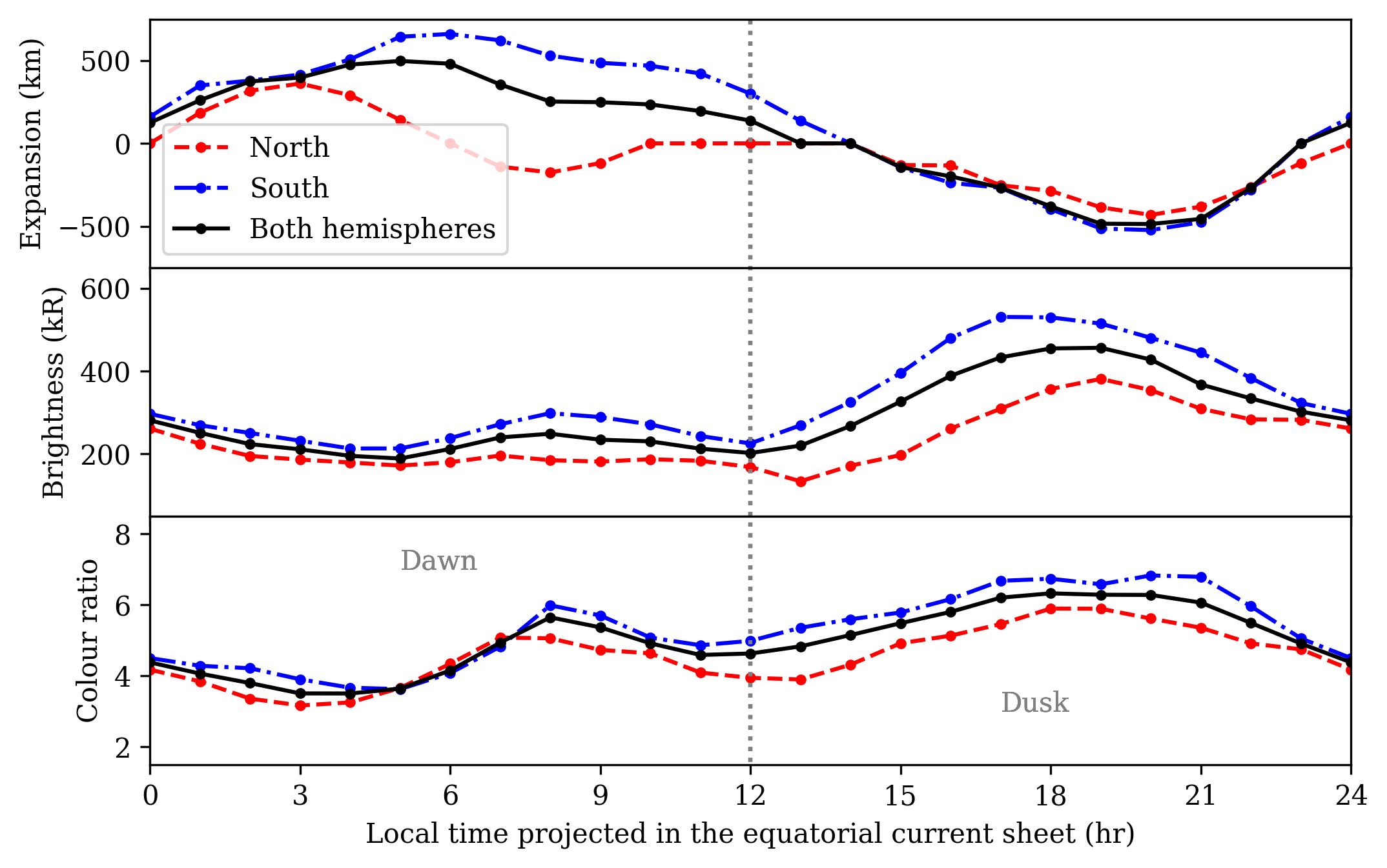}
    }
    \caption{
        The median-averaged expansion, characteristic brightness, and colour ratio of the ME as a function of projected local time in the ECS. The northern hemisphere is given by a red dashed line, the southern hemisphere by a blue dot-dashed line, and both hemispheres by a solid black line.   
    }
    \label{fig:local_time}
\end{figure}

In addition to investigating the global behaviour of the ME, it is equally possible to analyse how the properties of the ME change in local time, as shown in Fig. \ref{fig:local_time}.
The arcs of the ME were projected to the ECS using the JRM33 internal-field and Con2020 external-field models, as described in section \ref{section:mag_field_model}, to allow for more meaningful comparison between the northern and southern aurorae. 
The projected ME arcs were binned in ECS local time in three-hour bins separated by one hour, and then the median average of each bin taken; this overlap between bins serves to smooth the data shown in Fig. \ref{fig:local_time}. 

It can be seen in Fig. \ref{fig:local_time} that, in the north as well as the south, the ME tends to be contracted on the dusk side and expanded on the dawn side. 
This is in broad agreement with the local-time dependence of the ME mapping of \citet{vogt+:2011}, which was associated with a corotation-enforcement current or plasma outflow rate that is local-time dependent.
The northern hemisphere is typically less locally variable in expansion than the southern hemisphere, which appears to stand in contrast to the global variability discussed previously. 
However, one must be careful to distinguish between the local-time dependence of the expansion of the ME, which may be understood as a consequence of its local-time-dependent mapping to the ECS \citep{vogt+:2022}, and the global expansion of the ME relative to the reference oval, which has previously been linked to the effect of solar-wind pressure or the varying strength of the ECS magnetic field \citep{promfu+:2022}.
Here, it may be said that the expansion of the dawn-side northern ME depends slightly less strongly on local time than the conjugate ME in the south, though the comparatively poor coverage of UVS in the north (only 8 perijoves with northern ME coverages >80\%, against 25 in the south) limits the strength of this conclusion.

The characteristic brightness of the ME also shows a strong dependence on local time.
As in the discussion around Fig. \ref{fig:comp_vs_power}, the characteristic brightness is defined as the median-average brightness of all pixels in the detected arcs of the ME, now binned in local time.
The dusk-side ME is typically twice as bright as the dawn-side ME, with this difference being more pronounced in the south than in the north, which is in agreement with previous observations \citep{bonfond+:2015_1, groulard+:2024}.
These works noted that the power emitted from the dusk-side ME is around four times greater than that from the dawn-side ME. 
This is consistent with the result presented here, since emitted power is a consequence of both ME brightness and ME width, and the dusk-side ME is known to be around twice as wide as the dawn-side ME \citep{grodent+:2003}.
The simulations performed by \citet{chane+:2017} also predicted this dawn-dusk ME brightness asymmetry.
They also predicted a far greater day-night asymmetry in ME brightness; however, the dawn-dusk asymmetry in brightness in Fig. \ref{fig:local_time} is more striking than a day-night asymmetry, if one is indeed present. 
In contrast to the dusk-side ME, neither the day- nor night-side ME show consistently disrupted morphologies, so this lack of obvious day-night asymmetry is unlikely to be a result of poor sampling by the arc-detection algorithm.
The lack of obvious day-night asymmetry in the brightness of the ME is difficult to reconcile with the modelled \citep{chane+:2017} and observed \citep{lorch+:2020} predominance of FACs in the night-side magnetosphere, if FACs give rise to the ME, as is the case in the explanation related to corotation-enforcement; this is discussed in more depth in section \ref{sec:interpreting_in_context}. 
The colour ratio of the ME also peaks at dusk; this is likely a consequence of the established dependence of ME colour ratio on ME brightness \citep{gerard+:2016}.
The brightness and, more clearly, the colour ratio also show a secondary peak around 08:00.
This may be attributed to the presence of dawn storms, transient auroral features that appear on the dawn-side ME that show increased brightnesses and colour ratios \citep{bonfond+:2021}.
Dawn storms typically disrupt the clear arc-like nature of the ME, and thus are not included in the detected ME arcs.
If they were included, both the brightness and colour ratio would show strong peaks between 06:00 and 12:00, since dawn storms are bright, high-colour-ratio features that appears in around one-third of the UVS images used in this work, in line with previous estimates \citep{bonfond+:2021}.
However, it is possible that smaller, less disruptive pseudo-dawn storms \citep{bonfond+:2021} are included in the dataset, which are the likely origin of this secondary peak. 
These are small features that lie on the ME and so are unlikely to affect the measure of ME expansion, both locally and globally.

\subsection{Comparison with the Ganymede footprint}
Expansion or contraction of the ME can be potentially understood as the consequence of two physical processes. 
Firstly, the magnetic-field-line mapping between the ECS and the ionosphere may be variable.
For a fixed ME source region in the ECS, a variable magnetic mapping would move the ME poleward (contraction) if the field lines themselves were compressed, and equatorward (expansion) if the field lines were expanded. 
Alternatively, it may be that the ME source region itself varies in position in the magnetosphere, which would translate to a variable global ME expansion.
It is possible to use the position of the Ganymede footprint (GFP), as the moon closest to the presumed ME source region at 30 R$_{J}$ with a consistently visible auroral footprint \citep{hue+:2023}, to distinguish between the relative contributions of these two parameters \citep{vogt+:2022}.
Since the magnetospheric source region of the GFP, Ganymede, is unlikely to vary its distance from Jupiter as a function of e.g. magnetospheric compression, a correlation between the global ME expansion and the movement of the GFP relative to a fixed reference path would imply that a change in the magnetic mapping is largely responsible for the variable expansion of the ME. 
If the two show little or no correlation, it is likely that the movement of the ME magnetospheric source region plays a larger role.
Additionally, since the effect of a variable ECS magnetic field on the mapping of auroral features would become more prominent with distance into the magnetosphere, the latitudinal shift of the GFP is expected to correlate more strongly than the shift of the Europa footprint (EFP) with the ME expansion, which itself would correlate more strongly than the Io footprint (IFP).

In this work, the position of the IFP, EFP, and GFP were (where visible) manually identified in each of the UVS master images. 
The position of these spots were compared with the magnetic-field-line mapping (as per section \ref{section:mag_field_model}) of the orbit of their respective moons to an altitude of 900 km \citep{hue+:2023} from the 1-bar level to determine the shift of the moon footprint from a fixed reference path, in much the same way as the UVS ME reference was used in section \ref{section:uvs_ref_oval}.
Here, the use of the magnetically mapped moon-footprint path is preferred over the use of the latest empirical paths based on UVS observations \citep{hue+:2023} since it is precisely the shift in position of the footprints in UVS images that is being measured in this work, and so comparison with a path independent of UVS image data is more sensible.
The moon footprints are, in fact, made of multiple discrete spots \citep{bonfond+:2013_io,bonfond+:2013_gany}, but only the brightest of these spots, which is frequently but not always the Main-Alfv\'{e}n-Wing (MAW) spot \citep{hue+:2022}, was considered in this work. 
Since only the shift in magnetic latitudes of the footprints (perpendicular to the footprint path) are considered in this work, rather than any shift in longitude (along the footprint path), there is no need to distinguish between the various spots that compose the moon footprints. 
Indeed, previous work has noted that the latitudinal positions of the MAW spot and the footprint tail do not show any meaningful deviation \citep{moirano+:2024}, and so the position of the footprint tail is a suitable measure of the latitudinal shift of the moon footprint.  

\begin{figure}
    \centering
    \ifthenelse{\equal{\twocol}{"y"}}{
        \includegraphics[width=\linewidth]{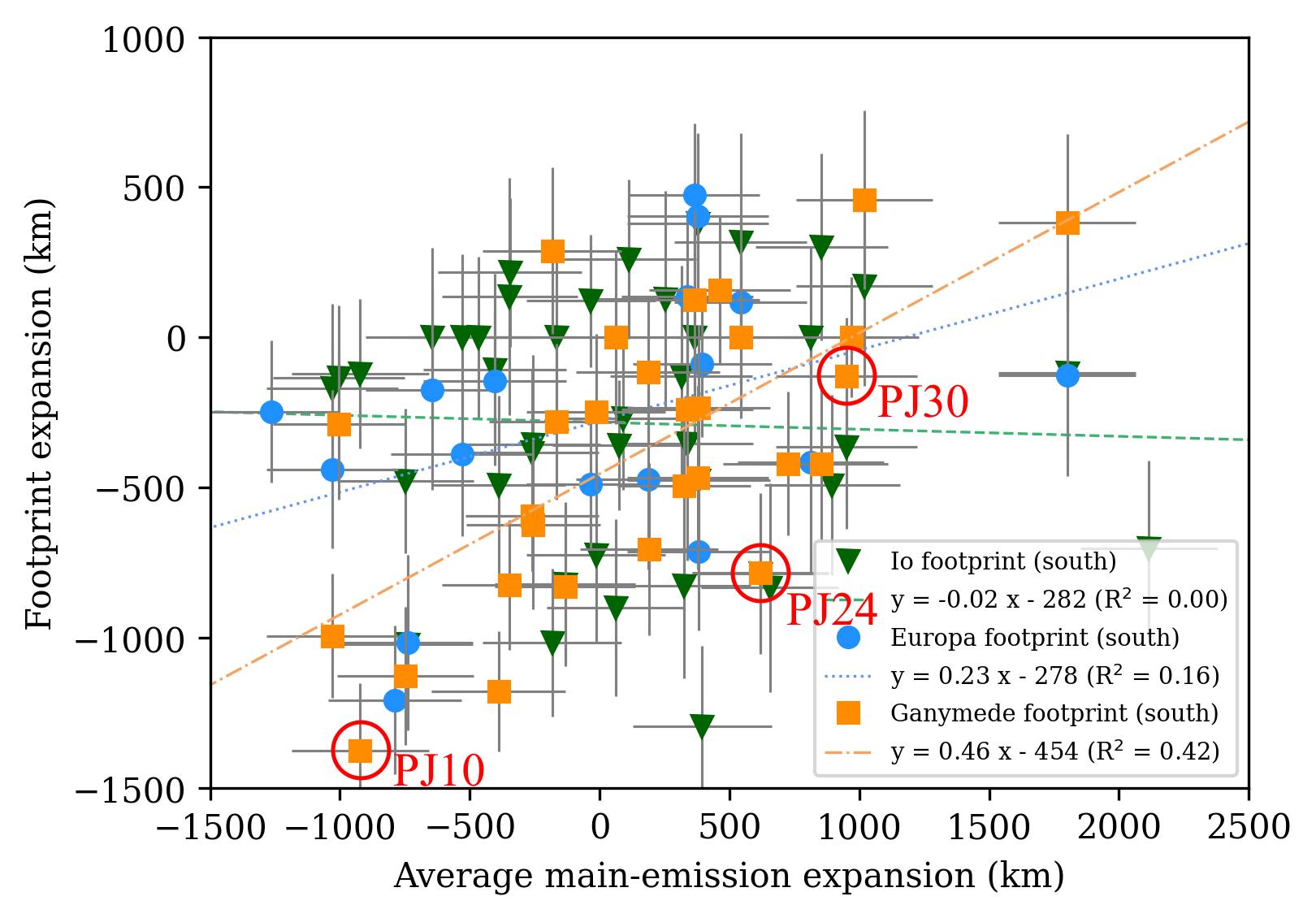}
    }
    {
        \includegraphics[width=\linewidth]{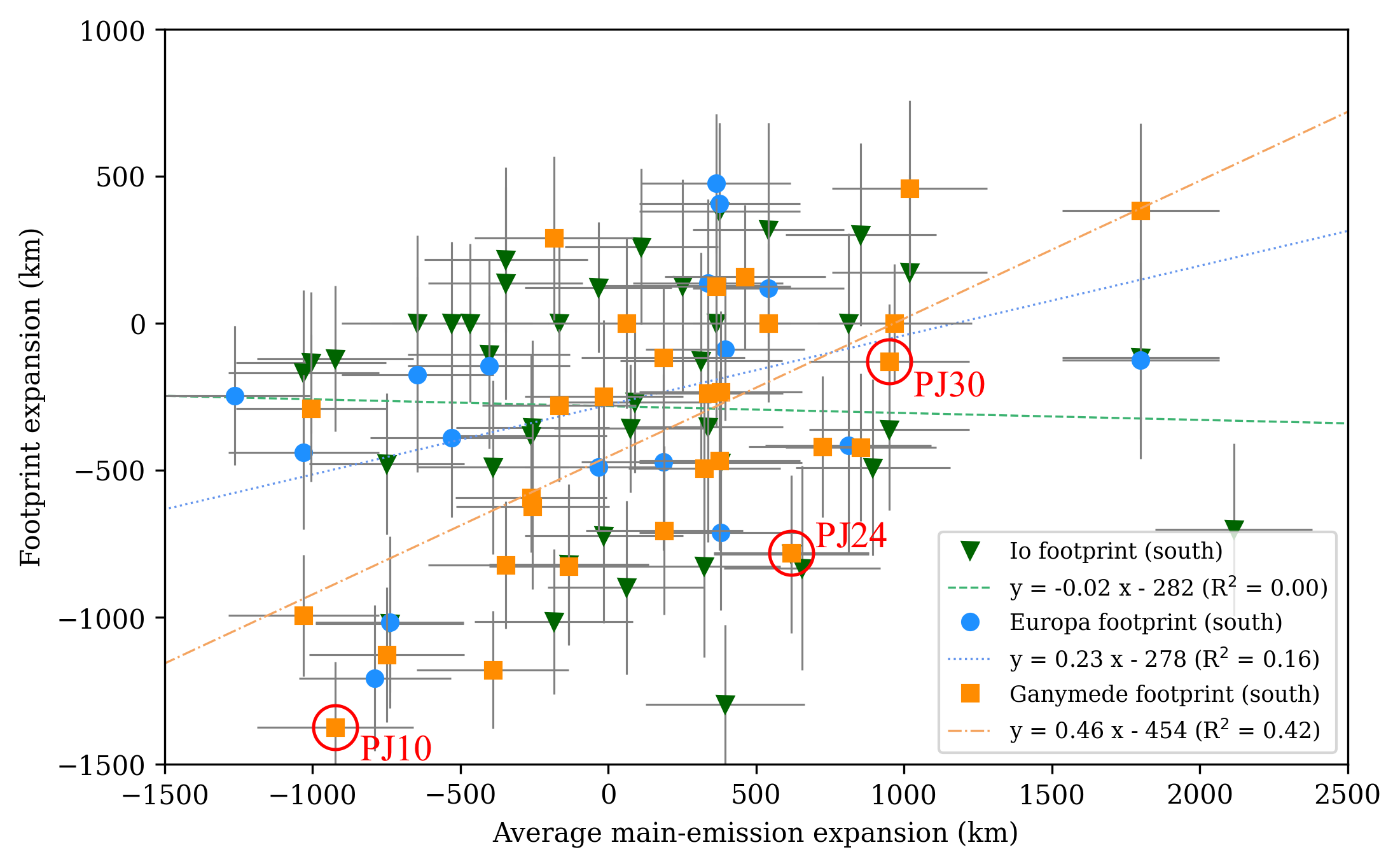}
    }
    \caption{
        The median-averaged expansion of the southern ME relative to the UVS reference oval vs the expansion of the Io, Europa, and Ganymede auroral footprints relative to their magnetically mapped contours at 900 km. The expansion of the Io footprint is denoted by green triangles, that of the Europa footprint by blue circles, and that of the Ganymede footprint by orange squares. The uncertainty in the expansion of the ME and the position of the moon footprints (estimated from their average apparent size as 3 px) is given for each point. The fitted relation between the Io-footprint expansion and the ME expansion is denoted by a green dashed line, that between the Europa-footprint location and the ME expansion by a blue dotted line, and that between the Ganymede-footprint location and the ME expansion by an orange dot-dashed line. The forms and R-squared goodness-of-fit coefficients of these relations are given in the legend. Cases at similar System-III longitudes have been annotated and highlighted with red circles.  
    }
    \label{fig:comp_vs_gfp}
\end{figure}

As shown in Fig. \ref{fig:comp_vs_gfp}, the latitudinal shift of the southern GFP relative to the magnetically mapped reference contour shows a reasonable (R$^2$ = 0.42) correlation with the ME expansion, in agreement with early results based on limited data \citep{grodent+:2008}. 
It is also in agreement with more recent work which found a moderate agreement between expansion of the day-side ME and latitudinal shift of the GFP \citep{vogt+:2022}, though the relation found in this work is both stronger and applicable over a wider range of ME expansions.
This result indicates that it is likely to be a changing magnetic mapping between the ME source region and the ionosphere that best explains the variability in the global expansion of the ME.
This may be due to a changing current intensity in the magnetodisc, and hence a changing contribution to the total magnetic field of Jupiter, which works to stretch the magnetic field lines outward.
This would correspond to an expansion of the ME when the magnetodisc current intensity is elevated, which can occur during periods of increased plasma outflow from the Io torus \citep{nichols:2011}. 
The magnetic-field mapping to the ionosphere from the ME source region and Ganymede, both at a greater distance from Jupiter than Io, would be influenced by the changing ECS magnetic field.
However, the mapping between the ionosphere and Io itself, where the magnetic field is essentially dipolar \citep{promfu+:2022}, would remain relatively unaffected; indeed, Fig. \ref{fig:comp_vs_gfp} shows that the footprint of Io does not demonstrate any correlation with the expansion of the ME (R$^2$ = 0.00).
The latitudinal shift of the EFP, as expected of an intermediary moon, shows a correlation with the ME expansion that has both a gradient and R$^{2}$ goodness-of-fit between those of the IFP and the GFP.
This indicates that the effect of the variable field-line stretching on the latitudinal shift of auroral features becomes more prominent with distance from Jupiter, as expected.
The majority of the cases in Fig. \ref{fig:comp_vs_gfp} show GFP latitudinal shifts in line with predictions from models ($\pm$650 km) \citep{moirano+:2024}.
Nevertheless, the IFP does show variations from its reference contour of comparable magnitude to those of the GFP.
This may be in part due to the weaker magnetic field in the ionosphere at lower latitudes, which amplifies even the smaller magnetic variations expected at Io to be of similar magnitude to those of the higher-latitude GFP in absolute (km) terms.
Typically, the ME moves twice as much compared to its reference contour than the GFP.
This is as expected, since the ME source region is around twice as far into the magnetosphere as Ganymede, where Jupiter's internal magnetic field is weaker, and thus the ME magnetic mapping is affected to a greater extent by an increased magnetodisc field than that of the GFP. 

There exists a considerable scatter in the GFP data around their fitted relation, which is possibly due, in part, to systematic under- or over-estimation of the average GFP latitude by parts of the magnetically mapped reference contour.
Given the comparatively few detections of the GFP in this dataset, it is not possible to robustly determine which parts of the reference contour show systematic inaccuracies and where the GFP shows a genuine deviation from its nominal location.
Nevertheless, the variation of the latitudinal shift of the GFP within a small region of the reference contour can be analysed in an effort to quantify this systematic error.
In Fig. \ref{fig:comp_vs_gfp}, three cases with similar GFP System-III longitudes in a region of the reference contour that is suspected to underestimate the latitude of the GFP have been highlighted: PJ10 ($\phi_{S3}$ = 272\textdegree), PJ24 ($\phi_{S3}$ = 283\textdegree), and PJ30 ($\phi_{S3}$ = 267\textdegree).
All three cases appear to show poleward GFP shifts despite the considerable range in ME expansion that they encompass, which indicates that the reference contour is indeed underestimating the latitude of the GFP in this range.
Additionally, these three cases show the same positive relation between the expansion of the ME and the latitudinal shift of the GFP, suggesting that some portion of the scatter around the fitted linear relation is due to systematic errors in the reference contour.

In all, the results indicate that a changing magnetic-field mapping, likely due to a variable contribution to the total magnetic field by the ECS, can largely account for the variable expansion of the ME.
Its R-squared value indicates that a linear relation between the expansions of the ME and the GFP can account for 46\% of the variability in the data.
Care must be taken, however, when using this goodness-of-fit value to make physical conclusions.
The conclusion offered, that a changing magnetic mapping largely accounts for the variable expansion of the ME, is not based simply on the relation between the expansion of the ME and the GFP, but rather on the combination of the linear relation between the expansion of the ME and the GFP, and the lack of linear relation between the expansion of the ME and the IFP.
In this, therefore, the statistical strength of this conclusion does not allow itself to be easily deduced from the goodness-of-fits of the two relations.
One can imagine the case where the two relations shown in Fig. \ref{fig:comp_vs_gfp} perfectly describe the data; in this case, both R-squared values would be unity, and the changing-magnetic-mapping model would be strongly supported.
However, if the relation between the ME and IFP expansions instead showed a positive gradient comparable with the relation between the ME and GFP expansions, the proposed conclusion would be less strongly supported, despite the R-squared values of unity, as it would no longer agree with the premise that the magnetic field at Io depends much less strongly on the state of the ECS.
Thus, the credibility of the proposed conclusion depends on a combination of the parameters of the linear relations, their R-squared values, and the accompanying physical interpretation.
This does not exclude other explanations for this variability, such as a moving ME source region, instead only indicating that the changing magnetic mapping implied by the relation shown in Fig. \ref{fig:comp_vs_gfp} accounts for a large portion of the variability in the data.
Indeed, in one set of HST images, the GFP was detected poleward of the ME \citep{bonfond+:2012}, which indicates that the ME source region can move.
This displacement of the ME source region was linked to an increased mass outflow rate from the orbit of Io, which would also work to stretch the magnetic field lines.
Thus, it is possible that this variability in position of the ME source region can partially account for the spread of the data in Fig. \ref{fig:comp_vs_gfp}.

\subsection{Comparison with magnetodisc current strength}

\begin{figure}
    \centering
    \ifthenelse{\equal{\twocol}{"y"}}{
        \captionsetup[subfigure]{width=0.95\linewidth}
        \subfloat[]{
            \includegraphics[width=\linewidth]{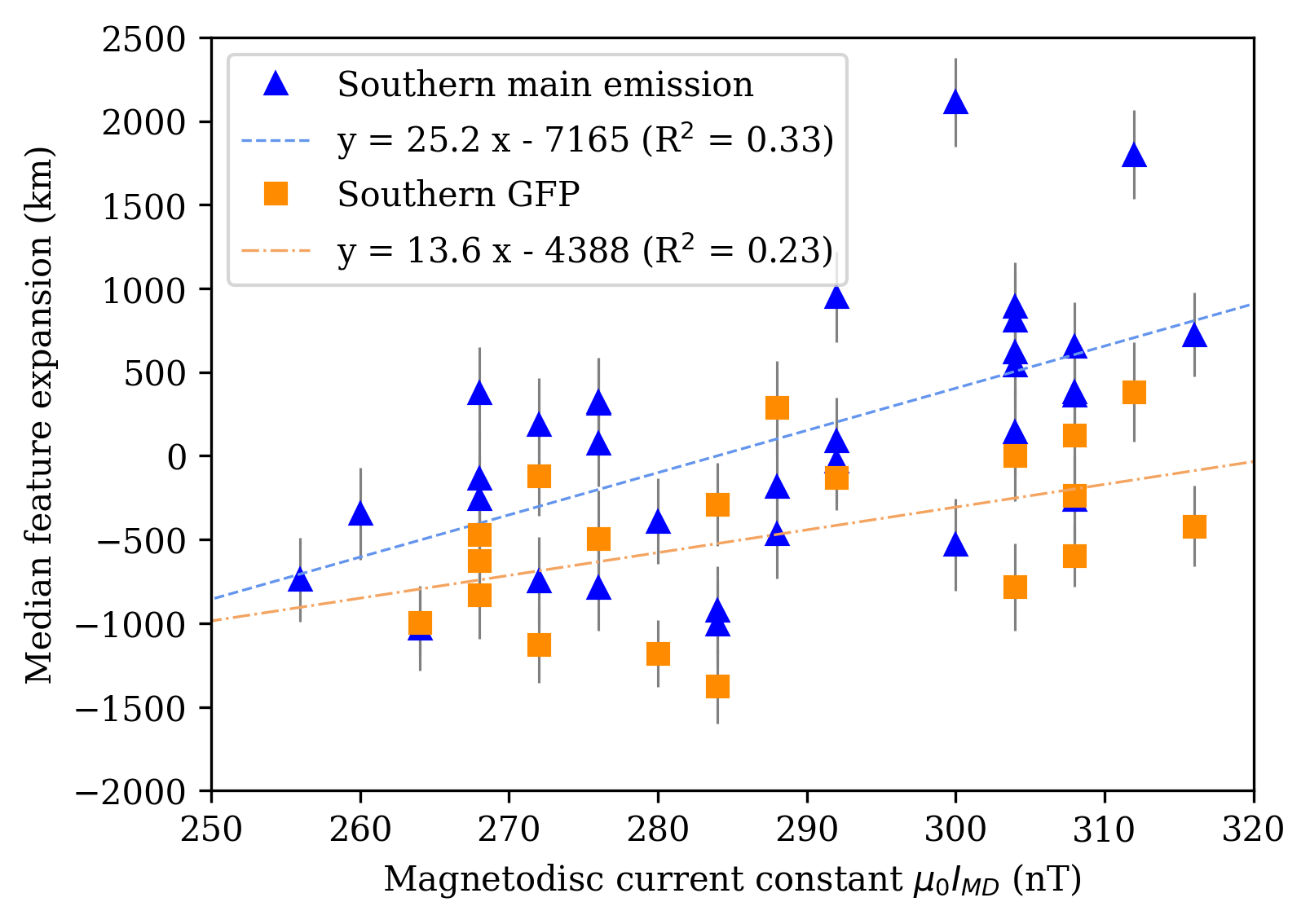}
        }\\
        \subfloat[]{
            \includegraphics[width=\linewidth]{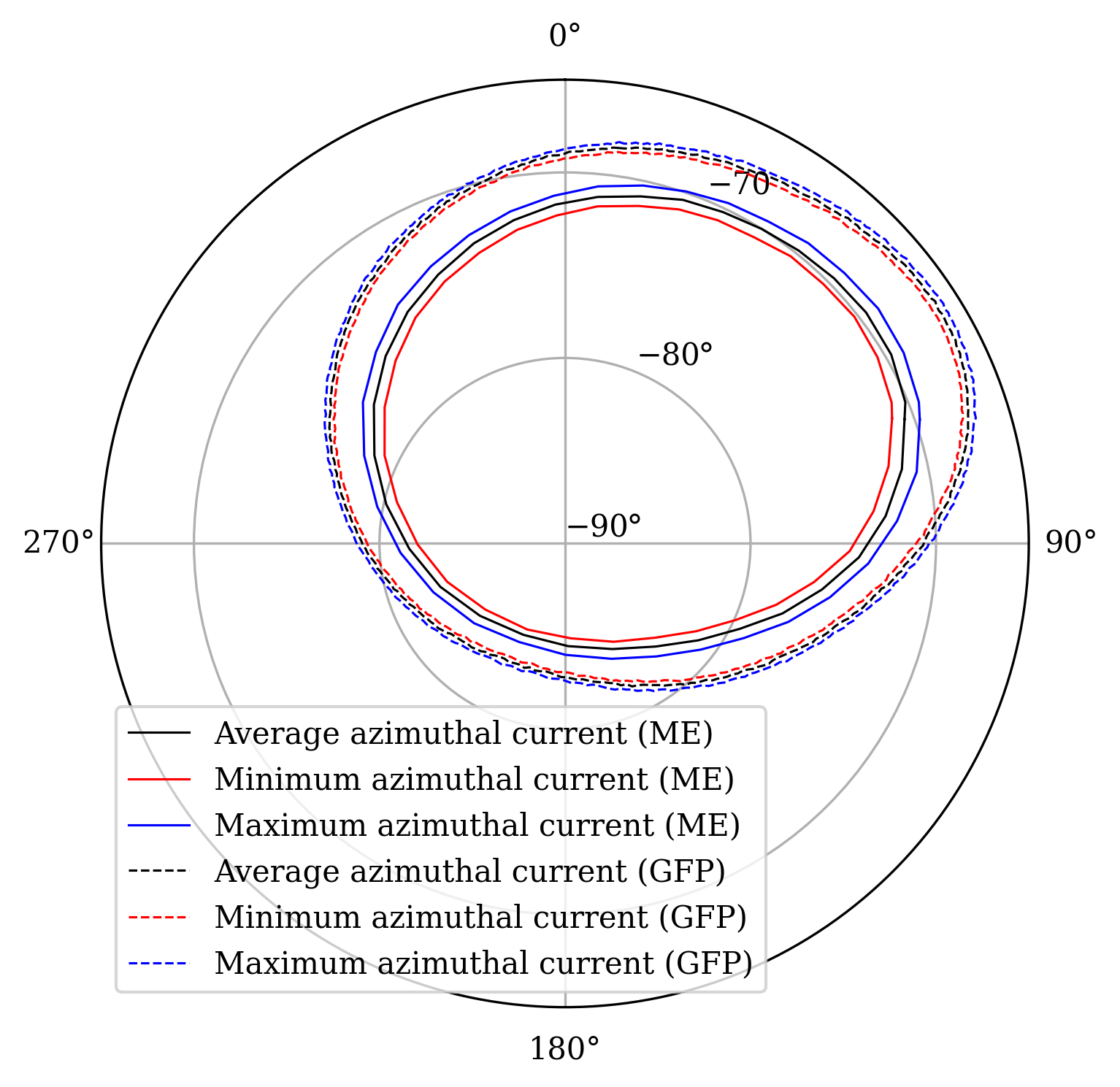}
        }
    }
    {
        \captionsetup[subfigure]{width=0.95\linewidth}
        \subfloat[]{
            \includegraphics[width=0.49\linewidth]{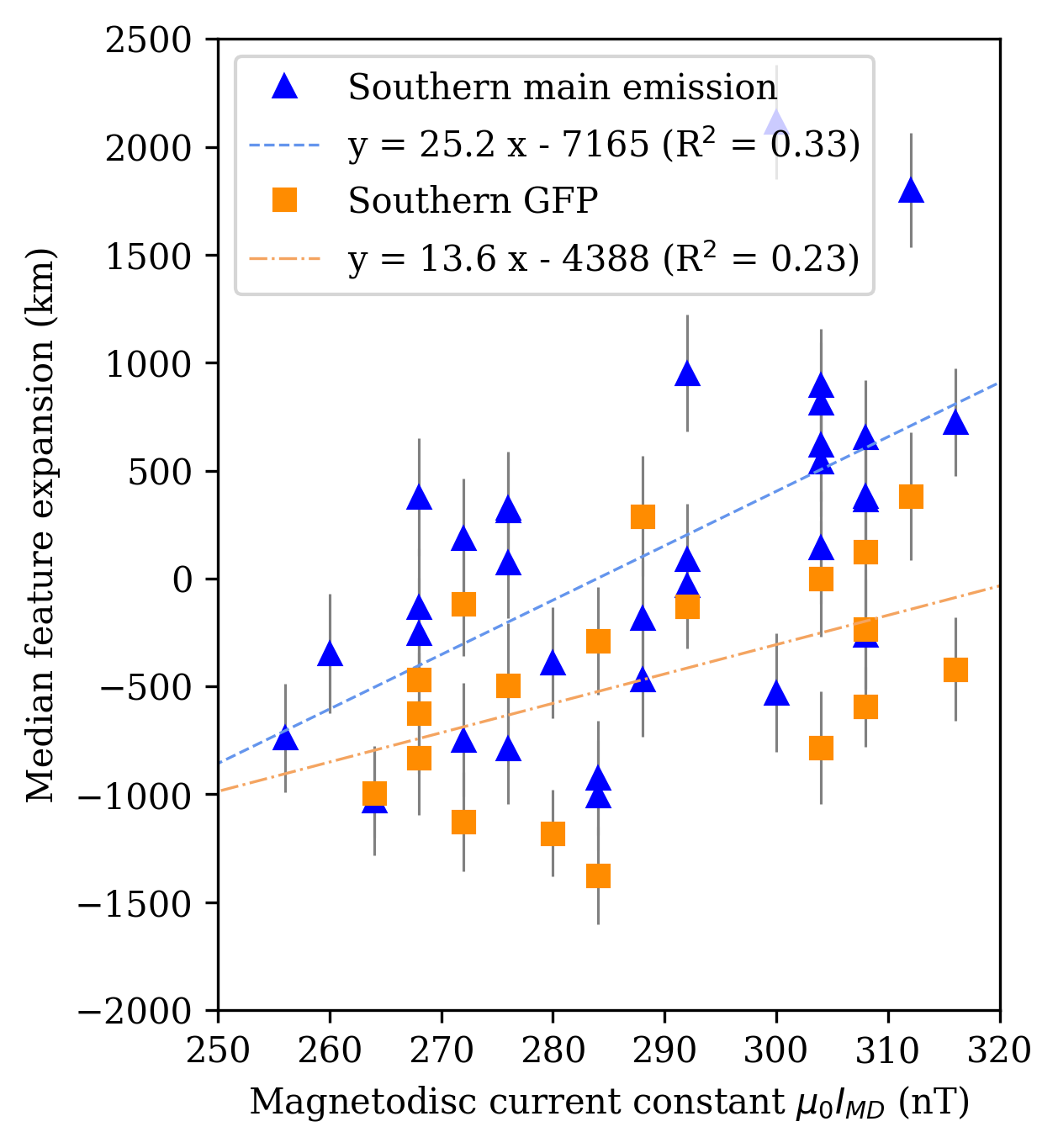}
        } 
        \subfloat[]{
            \includegraphics[width=0.49\linewidth]{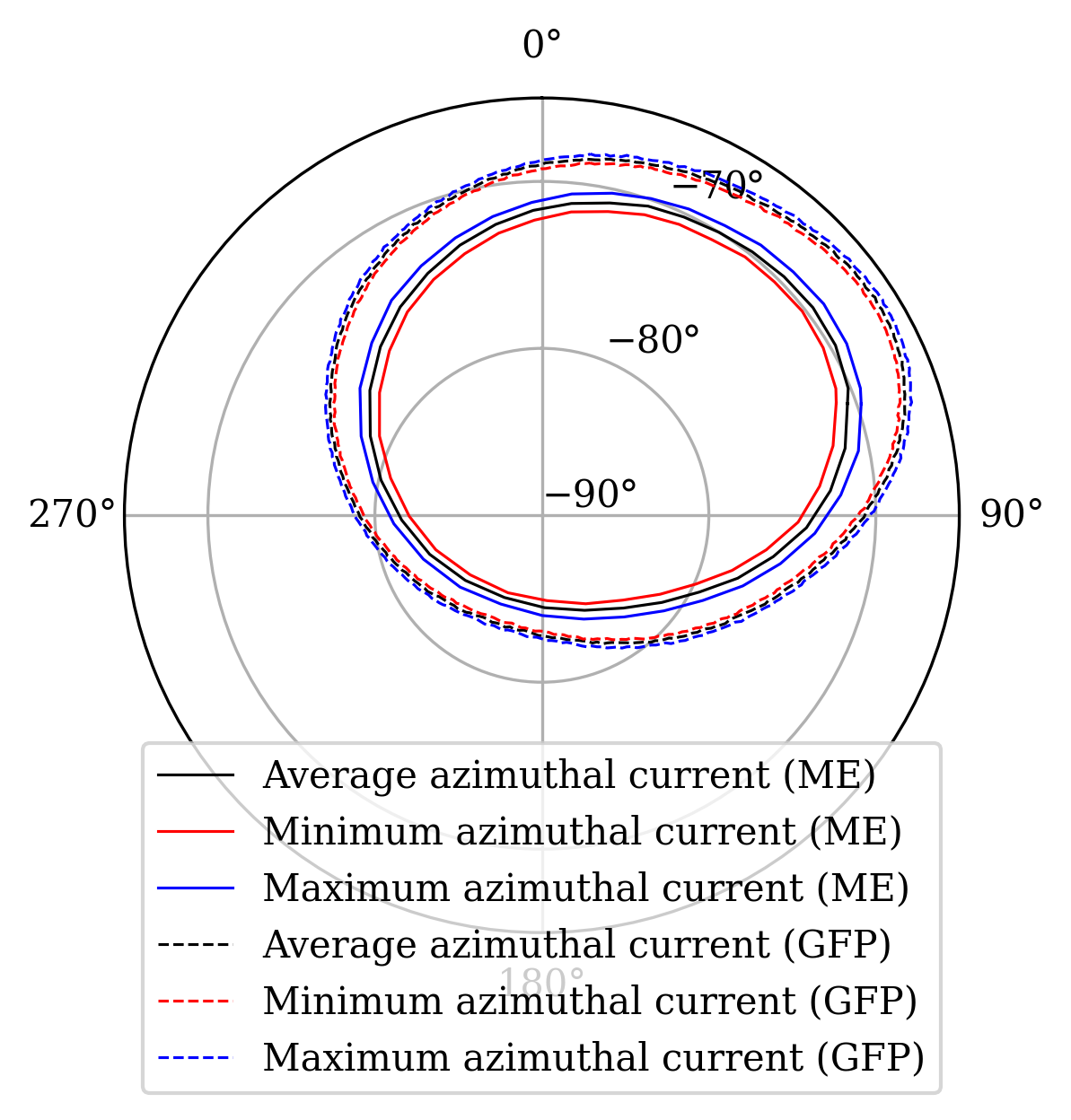}
        }
    }
    \caption{
        (a) The magnetodisc current constant fitted to perijoves 1 through 34 after \citet{vogt+:2022b} vs the median-averaged global expansion of the ME relative to the UVS reference oval and the expansion of the Ganymede auroral footprint relative to its mapped contour at 900 km, in the southern hemisphere. The global ME expansion in the south is denoted by blue triangles, and the expansion of the southern Ganymede footprint by orange squares. The fitted relation between the ME expansion and the current constant is denoted by a blue dashed line, and that between the Ganymede-footprint location and the current constant by an orange dot-dashed line. The form and R-squared goodness-of-fit values of the fitted relations are given in the legend.
        (b) The auroral mapping of the southern UVS ME reference oval (solid lines) and the southern GFP (dashed lines) using the JRM33 + Con2020 magnetic-field model with azimuthal current constants equivalent to the average (288 nT, black), minimum (256 nT, red, innermost), and maximum (316 nT, blue, outermost) values shown in (a). The System-III longitudes of gridlines are annotated around the outside of the plot, and the planetocentric latitudes next to circular gridlines.
    }
    \label{fig:conn2020_comp}
\end{figure}

Indeed, the relationship between the ECS magnetic-field strength and the expansion of the ME can be more directly studied using the magnetodisc current constants fitted to each perijove \citep{vogt+:2022b}, as shown in Fig. \ref{fig:conn2020_comp}a.
The global expansion of the ME shows a positive correlation with the magnetodisc current constant, which is consistent with an outward stretching of the magnetic field lines by the magnetodisc \citep{vogt+:2022} and hence with the conclusions drawn from Fig. \ref{fig:comp_vs_gfp}.
The same relationship is also present in the perpendicular shift in the GFP, as expected from Fig. \ref{fig:comp_vs_gfp}.
In both cases, a linear relation with the magnetodisc current constant accounts for around one third of the variation in the data.
This relationship was not previously found in a similar analysis of HST data \citep{vogt+:2022}, though this is possibly due to the large uncertainty in the limb fitting of Jupiter in HST images \citep{bonfond+:2017}.
\citet{vogt+:2022b} estimated an average magnetodisc current constant $\mu_0 I_{MD}$ of 288 nT; the fitted relationship for the southern ME in Fig. \ref{fig:conn2020_comp}a predicts a very small global ME expansion of 92 km (less than one pixel in the polar-projected images) at this value of magnetodisc current constant, which supports the use of the ME reference oval defined in this work as the average position of the ME.
This relationship can be interpreted in the context of an outward stretching of the global magnetic field when the azimuthal current in the magnetodisc, and hence the ECS contribution to the magnetic field, is elevated, which leads to the mapping of the fixed ME magnetospheric source region to a smaller M-shell and therefore an expansion of the ME \citep{vogt+:2017}.
Field-line tracing using the JRM33 + Con2020 magnetic-field model (Fig. \ref{fig:conn2020_comp}b) indicates that an increased magnetodisc current constant should indeed lead to an expansion of the ME.
The model predicts a ME contraction of -650 km at the minimum magnetodisc current constant and an expansion of 660 km at the maximum, which is in agreement with the fitted relation in Fig. \ref{fig:conn2020_comp}a to within the uncertainty of the data.
Fig. \ref{fig:conn2020_comp}b also shows that the GFP is expected to expand or contract with the ME but to a smaller absolute extent ($\pm$310 km vs $\pm$660 km for the ME), which is in quantitative agreement with the fitted relation shown in Fig. \ref{fig:comp_vs_gfp}.

This variability in azimuthal current may be the result of variable plasma mass outflow from the Io torus, which works to stretch the magnetic field lines outward and hence move the ME and GFP equatorward \citep{nichols+:2009b}.
Additionally, compression of the magnetosphere by the solar wind has been observed increase current-sheet intensity \citep{xu+:2023} and to move the GFP \citep{promfu+:2022} and the day-side ME (see section \ref{sec:ms_compression_state} below) poleward, which may explain some portion of the variance in the data shown in Fig. \ref{fig:conn2020_comp}.
Therefore, the variability in the expansion of the ME is likely a combination of both internal (mass loading from the Io torus) and external (solar wind) sources.
To distinguish between these two sources, information regarding the timescale of the change in the ME expansion is required. 
Changes due to solar-wind pressure are expected to occur over periods of several hours \citep{chane+:2017}, whereas those changes due to increased plasma outflow from the Io torus are expected to happen over longer timescales of several weeks \citep{bagenal+delamere:2011,nichols+:2017,tao+:2018}.

\subsection{Comparison with magnetospheric compression state}
\label{sec:ms_compression_state}

\begin{figure}
    \centering
    \ifthenelse{\equal{\twocol}{"y"}}{
        \includegraphics[width=\linewidth]{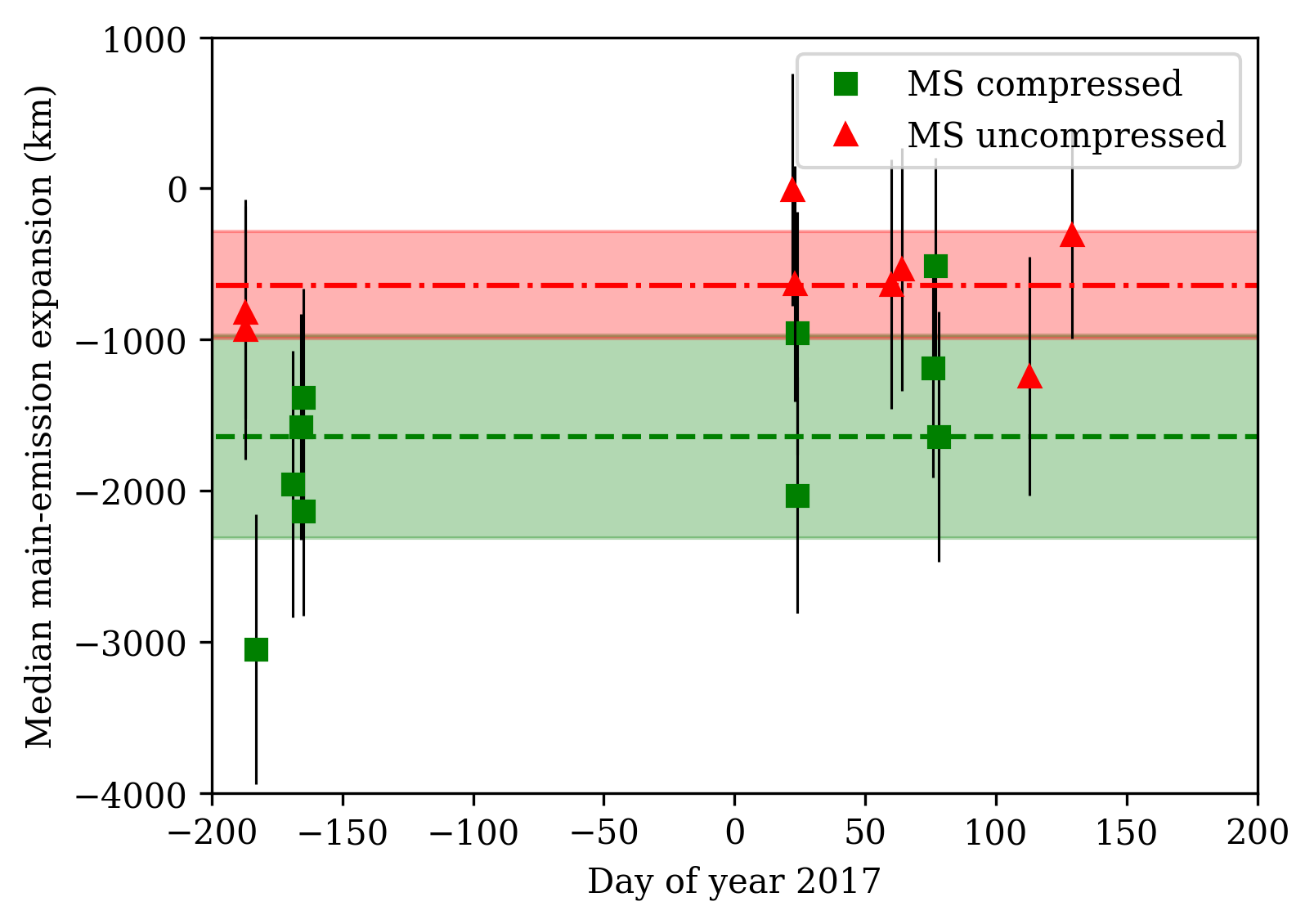}
    }
    {
        \includegraphics[width=\linewidth]{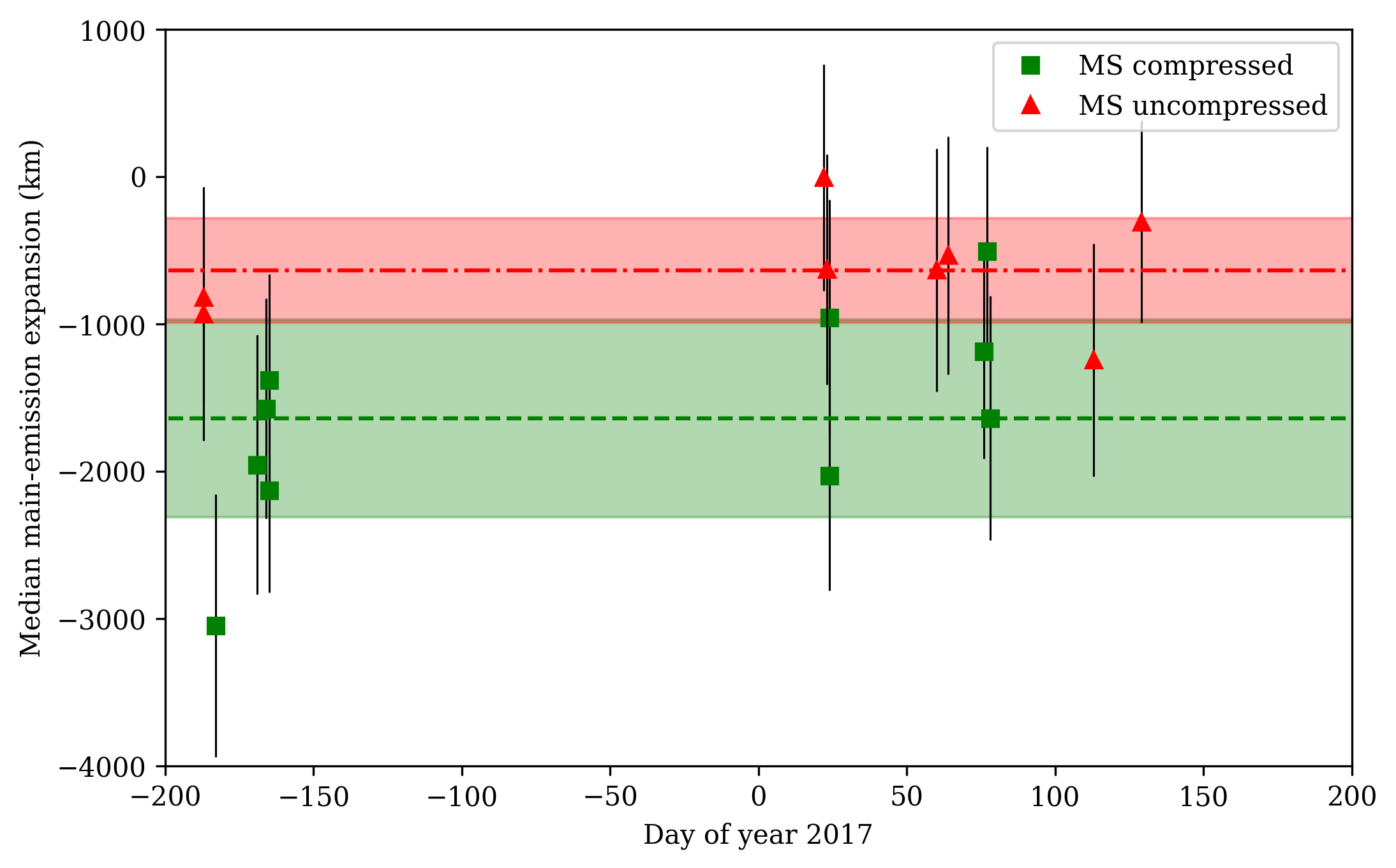}
    }
    \caption{
        The imaging date (as day-of-year 2017) vs the median-averaged global ME expansion from the UVS reference oval for those northern-hemisphere HST image series with known magnetospheric (MS) compression states, after \citet{yao+:2022}. An uncompressed magnetosphere is denoted by a red square and a compressed magnetosphere by a green triangle. The average ME expansion in the uncompressed-magnetosphere case (-600 km) is denoted by a red dot-dashed line, and that in the compressed-magnetosphere case (-1600 km) by a green dashed line. The shaded regions around each average-value line denote the 1$\sigma$ range.
    }
    \label{fig:ms_compression}
\end{figure}

As Juno passes through the magnetopause, it is possible to determine the state of compression of the magnetosphere \citep{yao+:2022}.
This is accomplished via the detection of trapped low-frequency radio continuum radiation, which indicates the crossing of the magnetosheath by Juno; were the magnetosphere compressed, this crossing would occur at a lower altitude. 
Since Juno is necessarily far from the planet when this occurs, no UVS image data are collected for comparison with the compression state of the magnetosphere.
However, in many cases, HST image data are available during the period that Juno crosses the magnetopause. 
These HST data come with two caveats: firstly, that only the day-side ME can be imaged by HST; and secondly, that HST-STIS images have relatively large uncertainties in the centring of Jupiter ($\pm$8 px $\sim$ 800 km) \citep{bonfond+:2017}.
Additionally, only northern-hemisphere cases have been considered due to their favourable viewing geometry and a paucity of suitable HST image series imaging the southern hemisphere.

As shown in Fig. \ref{fig:ms_compression}, the average global ME contraction in the northern hemisphere shows a strong correlation with the compression state of the magnetosphere; when the magnetosphere is compressed, the day-side ME is also contracted.
This behaviour is in line with results from modelling work, which indicate that increased solar-wind dynamic pressure compresses the day-side field lines, moving the day-side ME poleward, and that increased plasma mass outflow from the Io torus stretches the day-side magnetic field lines outwards, moving the day-side ME equatorward \citep{promfu+:2022}.
The 1$\sigma$ ranges of the two data series overlap only very slightly, indicating that the difference in the average contraction of the ME in magnetosphere-compressed and magnetosphere-uncompressed cases is of reasonable statistical significance.  
It should be noted here that, due to the typically more disrupted morphology of the dusk-side ME during magnetospheric compression \citep{yao+:2022} and its resulting unsuitability for the arc-detection algorithm, the compressed-magnetosphere cases in Fig. \ref{fig:ms_compression} are likely biased toward the dawn-side ME.
However, this is not necessarily an issue when interpreting Fig. \ref{fig:ms_compression}.
It has been previously shown (Fig. \ref{fig:local_time}) that the dawn-side ME tends to undergo expansion from the average ME position, and that, conversely, the dusk-side ME tends to be contracted; did the dusk-side ME not show this disruption during magnetospheric compression, and were it hence more consistently included when determining the global ME expansion, it would likely only serve to make the distinction between magnetosphere-compressed and magnetosphere-uncompressed cases more striking.
Additionally, in HST images of the northern hemisphere, the dusk-side ME is located in the region of the low-strength magnetic anomaly and therefore moves more in kilometre terms than the dawn-side ME for a given change in the magnetic field, which would lead to an even greater distinction between the magnetosphere-compressed and magnetosphere-uncompressed cases if the dusk-side ME was more consistently included.

\subsection{Interpreting the results in the context of theories of ME generation}
\label{sec:interpreting_in_context}

The FACs that are responsible for generating the ME in the corotation-enforcement-current model \citep{cowley+bunce:2001} have been modelled \citep{chane+:2017,sarkango+:2019} and observed \citep{lorch+:2020} to be stronger in the night-side magnetosphere than in the day-side magnetosphere.
If these currents give rise to the ME, it would be expected that the ME be brighter at night than during the day.
Additionally, multiple models \citep{chane+:2017,feng+:2022,sarkango+:2019} of the magnetospheric current system predict, under conditions of magnetospheric compression by the solar wind, an increase in the night-side FACs, which would work to increase the expected day-night asymmetry in the brightness of the ME.

These predictions are not supported by the results of this work.
The dominant asymmetry in the brightness of the ME is a dawn-dusk asymmetry (Fig. \ref{fig:local_time}); the dusk-side ME is observed to be around three times as bright as the dawn-side ME.
A day-night symmetry, if present, is far less striking than this dawn-dusk asymmetry.
This does not agree with the modelled and observed distribution of FACs in the jovian magnetosphere.
To link the contraction of the ME with compression of the magnetosphere, several steps are required.
Firstly, it has been determined that the day-side ME and the night-side ME expand and contract together (Fig. \ref{fig:night_day_comparison}); under conditions of global contraction of the ME, both the day- and night-side are also contracted from their average location.
Secondly, under conditions of magnetospheric compression, the day-side ME was observed to be consistently contracted (Fig. \ref{fig:ms_compression}).
These two results suggest that compression of the magnetosphere results in a global contraction of the ME. 
Under global contraction, the ME is observed to undergo global brightening.
This brightening is both more prominent and better correlated with ME contraction in the day-side ME (Fig. \ref{fig:comp_vs_power}).
The brightness of the night-side ME is both less prominent and shows a poorer (though still present) correlation with global ME contraction than the brightness of the ME as a whole.
This is in disagreement with the modelled response of the FACs to solar-wind compression, in which the night-side FACs more greatly increase in strength than the day-side FACs \citep{chane+:2017}; the model of \citet{sarkango+:2019} even predicts a diminishing in the day-side FACs during solar-wind compression.

An Alfv\'{e}nic model for the generation of the ME may better explain these results.
It has been previously estimated that the Alfv\'{e}nic Poynting flux is of the order of 62 to 620 mW m$^{-2}$ in the auroral acceleration region \citep{pan+:2021}, which is consistent with the downward energy fluxes, thought to give rise to the jovian aurorae, measured by the JEDI instrument aboard Juno \citep{mauk+:2017}.  
The dusk-side middle magnetosphere is known, from Galileo magnetometer measurements, to have a greater degree of turbulence than the dawn-side middle magnetosphere; under the Alfv\'{e}nic framework of \citet{saur+:2003}, this would correspond to a greater generation of Alfv\'{e}n waves in the dusk-side magnetosphere and hence a brighter dusk-side ME, as demonstrated in this work. 
During periods of compression of the magnetosphere by the solar wind, the model of \citet{feng+:2022} predicts an increase in auroral Alf\'{e}nic power, most notably in the day-side aurora, which is broadly consistent with the findings of this work.
However, in their simulation, this increase in day-side Alfv\'{e}nic power is also accompanied by an increase in the day-side FACs, and so this model does not necessarily support an Alfv\'{e}nic framework over a FAC framework.
The peak in the intensity of the FACs does not correspond to exactly the same location in the aurora nor does it occur at exactly the same time after the solar-wind shock as the peak in the Alfv\'{e}nic Poynting flux. 
Additionally, this model would indicate that, under uncompressed magnetospheric conditions, the aurora is brightest in the day- and dawn-side sectors, which is not consistent with other models nor observations. 
Solar-wind compression of the magnetosphere also leads to an expansion of the ME in this model, which is again inconsistent with the results of this work.
As it stands, neither the proposed FAC-based nor Alfv\'{e}nic ME-generation mechanisms are fully consistent with observation, and deeper analysis of turbulence within the magnetosphere is required. 
Additionally, information regarding the timescale of the changes in the expansion of the ME is necessary to distinguish between the response of the ME to solar-wind compression and torus-mass-outflow inflation of the magnetosphere.

\section{Conclusions}

The findings of this work can be summarised as such:
\begin{enumerate}
    \item In Juno-UVS image data between perijoves 1 and 54, Jupiter's main auroral emission was observed to globally expand and contract by as much as $\pm$2000 km from its average position. There is excellent correlation between the expansion in the northern and southern hemispheres, which indicates that the process(es) causing this expansion or contraction are global within the magnetosphere, as well as between the expansion of the day-side and night-side ME, suggesting that the processes that work to contract the ME affect both hemispheres simultaneously and are stable over timescales of several hours.
    \item The global expansion of the ME is anti-correlated with its brightness in both the northern and southern hemispheres; a contracted ME is typically brighter than an expanded ME. This brightening is more pronounced in the day-side ME.
    \item Additionally, the local morphology of the ME is asymmetric in local time; the dawn-side ME is typically expanded and the dusk-side ME typically contracted compared to the average ME position.
    \item The perpendicular shift of the auroral footprint of Ganymede from its magnetically mapped reference path is positively correlated with the global expansion of the ME, while the shift of the IFP from its reference position shows no correlation, which indicates that a variable magnetodisc magnetic field can account for a considerable part of the variability of the expansion of the ME. The behaviour of the EFP was found to be intermediate to that of the IFP and GFP, which is consistent with this interpretation.  
    \item The equatorward expansion of the ME for perijoves 1 to 20 correlates well with increased magnetodisc current constant, reinforcing the previous conclusion that the current-sheet magnetic field is an important factor in determining the expansion of the ME. 
    \item An analysis of the day-side expansion of the ME in HST images of the aurora showed a clear distinction in global expansion between those cases with compressed and uncompressed magnetospheres. When combined with the correlation between the day-side and night-side expansion of the ME, this indicates that an increased compression of the magnetosphere works to compress the magnetic field lines and hence move the ME poleward.
    \item The combination of these results suggests that solar-wind compression of the jovian magnetosphere works to increase the global brightness of the ME, though predominantly that of the day-side ME. This result stands in opposition to models and observations of the field-aligned currents in the middle magnetosphere, which are expected to give rise to the ME in the corotation-enforcement-current framework.
\end{enumerate}

\section*{Data availability statement}
Juno-UVS data can be obtained from the NASA Planetary Data System (\url{https://pds-atmospheres.nmsu.edu/data_and_services/atmospheres_data/JUNO/juno.html}). 
This work uses data from HST campaigns GO-14105 and GO-14634 which can be accessed via the Space Telescope Science Institute, operated by AURA for NASA and accessible at \url{https://archive.stsci.edu/hst/}.

\section*{Acknowledgements}
   We are grateful to NASA and contributing institutions which have made the Juno mission possible. This work was funded by NASA's New Frontiers Program for Juno via contract with the Southwest Research Institute. This publication benefits from the support of the French Community of Belgium in the context of the FRIA Doctoral Grant awarded to L. A. Head. B. Bonfond is a Research Associate of the Fonds de la Recherche Scientifique - FNRS. M. F. Vogt was supported by NASA grants 80NSSC17K0777 and 80NSSC20K0559. V. Hue acknowledges support from the French government under the France 2030 investment plan, as part of the Initiative d’Excellence d’Aix-Marseille Universit\'{e} – A*MIDEX AMX-22-CPJ-04.

\bibliography{references}

\begin{appendix}






\renewcommand{\thefigure}{B.\arabic{figure}}
\renewcommand{\thetable}{A.\arabic{table}}
\onecolumn

\begin{landscape}
\begin{picture}(0,0)
\put(330,0){\rotatebox{-90}{\parbox{10cm}{\section{Supplementary tables}}}}
\end{picture}
\begin{ThreePartTable}
    \centering
    \setlength\LTleft{0pt}
    \setlength\LTright{0pt}
    \setlength\tabcolsep{4pt}


    \begin{TableNotes}
      \item\footnotesize ``Perijove'' gives both the perijove number and hemisphere for each case. ``Date'' gives the date and time of the central spin of the master UVS image for that case, in the format YYYY-MM-DD hh:mm:ss. ``$\phi_{SS}$'' denotes the left-handed System-III subsolar longitude of Jupiter during the central spin of the master image. ``Coverage'' refers to the pseudo-magnetic angular coverage of all arcs detected within the ME for a given case (and not the coverage of the aurora by UVS). ``Expansion'' refers to the global expansion of the ME relative to the reference contour. Expansions are given as positive where the detected ME is, on average, equatorward of the reference contour. Negative values indicate global contraction of the ME compared to the reference contour. Missing values are attributed to insufficient UVS coverage which made it impossible to provide sensible estimates of the global expansion of the ME. Columns of the type ``Fig. X'' state whether a particular case was used in the analysis (or the part of the analysis given in parantheses) associated with the numbered figure. During PJ2, the Juno spacecraft entered safe mode and no data were acquired. For PJ52, the coverage in both the northern and southern hemispheres was too poor to allow estimation of the expansion of the ME.
    \end{TableNotes}
   
    \begin{longtable}[htbp]{ lllll p{1cm} p{1cm} p{1cm} p{1cm} p{1cm} p{1cm} p{1cm} p{1cm} }
        \label{tab:uvs_cases}\\
        \caption{Juno-UVS cases used in this work.}\\
    
        \toprule
        Perijove & Date & 	$\phi_{SS}$ (\textdegree) & Coverage &		 Expansion (km) & Fig. \ref{fig:north_south_comparison} & Fig. \ref{fig:night_day_comparison} & Fig. \ref{fig:comp_vs_power} & Fig. \ref{fig:comp_vs_gfp} (IFP) & Fig. \ref{fig:comp_vs_gfp} (EFP) & Fig. \ref{fig:comp_vs_gfp} (GFP) & Fig. \ref{fig:conn2020_comp} (ME) & Fig. \ref{fig:conn2020_comp} (GFP) \\
        \midrule
        \endhead

        \midrule[\heavyrulewidth]
        \multicolumn{13}{r}{\textit{continued...}}\\
        \endfoot

        \midrule[\heavyrulewidth]
        \insertTableNotes  
        \endlastfoot
   
    
                

         PJ1  N & 2016-08-27  12:01:51 &154& 81\% & 100  $\pm$  300 &Yes&Yes& \textcolor{gray}{No} & \textcolor{gray}{No} & \textcolor{gray}{No} & \textcolor{gray}{No} & \textcolor{gray}{No} & \textcolor{gray}{No}  \\ 
 PJ1  S & 2016-08-27  14:52:06 &257& 66\% & 700  $\pm$  300 &Yes&Yes& \textcolor{gray}{No} &Yes& \textcolor{gray}{No} & \textcolor{gray}{No} &Yes& \textcolor{gray}{No}  \\ 
 PJ3  N & 2016-12-11  16:40:25 &72& 71\% & 1300  $\pm$  300 &Yes&Yes& \textcolor{gray}{No} & \textcolor{gray}{No} & \textcolor{gray}{No} & \textcolor{gray}{No} & \textcolor{gray}{No} & \textcolor{gray}{No}  \\ 
 PJ3  S & 2016-12-11  19:11:02 &163& 80\% & 300  $\pm$  300 &Yes&Yes&Yes&Yes& \textcolor{gray}{No} &Yes&Yes&Yes \\ 
 PJ4  N & 2017-02-02  12:42:45 &342& 74\% & -600  $\pm$  300 &Yes&Yes& \textcolor{gray}{No} & \textcolor{gray}{No} & \textcolor{gray}{No} & \textcolor{gray}{No} & \textcolor{gray}{No} & \textcolor{gray}{No}  \\ 
 PJ4  S & 2017-02-02  14:37:01 &51& 91\% & -100  $\pm$  300 &Yes&Yes&Yes&Yes& \textcolor{gray}{No} &Yes&Yes&Yes \\ 
 PJ5  N & 2017-03-27  08:45:48 &254& 69\% & -1700  $\pm$  300 &Yes&Yes& \textcolor{gray}{No} & \textcolor{gray}{No} & \textcolor{gray}{No} & \textcolor{gray}{No} & \textcolor{gray}{No} & \textcolor{gray}{No}  \\ 
 PJ5  S & 2017-03-27  11:38:04 &358& 69\% & -800  $\pm$  300 &Yes&Yes& \textcolor{gray}{No} & \textcolor{gray}{No} &Yes& \textcolor{gray}{No} &Yes& \textcolor{gray}{No}  \\ 
 PJ6  N & 2017-05-19  05:34:29 &192& 88\% & -400  $\pm$  300 &Yes&Yes& \textcolor{gray}{No} & \textcolor{gray}{No} & \textcolor{gray}{No} & \textcolor{gray}{No} & \textcolor{gray}{No} & \textcolor{gray}{No}  \\ 
 PJ6  S & 2017-05-19  07:40:01 &268& 77\% & -300  $\pm$  300 &Yes&Yes& \textcolor{gray}{No} &Yes& \textcolor{gray}{No} &Yes&Yes&Yes \\ 
 PJ7  N & 2017-07-11  01:05:59 &84& 85\% & -400  $\pm$  300 &Yes&Yes& \textcolor{gray}{No} & \textcolor{gray}{No} & \textcolor{gray}{No} & \textcolor{gray}{No} & \textcolor{gray}{No} & \textcolor{gray}{No}  \\ 
 PJ7  S & 2017-07-11  03:51:29 &184& 97\% & 0  $\pm$  200 &Yes&Yes&Yes&Yes&Yes& \textcolor{gray}{No} &Yes& \textcolor{gray}{No}  \\ 
 PJ8  N & 2017-09-01  21:20:23 &2& 59\% & 600  $\pm$  300 &Yes&Yes& \textcolor{gray}{No} & \textcolor{gray}{No} & \textcolor{gray}{No} & \textcolor{gray}{No} & \textcolor{gray}{No} & \textcolor{gray}{No}  \\ 
 PJ8  S & 2017-09-01  23:28:24 &80& 89\% & 400  $\pm$  300 &Yes&Yes&Yes&Yes&Yes&Yes&Yes&Yes \\ 
 PJ9  N & 2017-10-24  17:23:42 &274& 52\% & -600  $\pm$  300 &Yes&Yes& \textcolor{gray}{No} & \textcolor{gray}{No} & \textcolor{gray}{No} & \textcolor{gray}{No} & \textcolor{gray}{No} & \textcolor{gray}{No}  \\ 
 PJ9  S & 2017-10-24  19:22:38 &346& 62\% & -300  $\pm$  300 &Yes&Yes& \textcolor{gray}{No} &Yes& \textcolor{gray}{No} &Yes&Yes&Yes \\ 
 PJ10  N & 2017-12-16  17:55:38 &347& 59\% & -2100  $\pm$  300 &Yes&Yes& \textcolor{gray}{No} & \textcolor{gray}{No} & \textcolor{gray}{No} & \textcolor{gray}{No} & \textcolor{gray}{No} & \textcolor{gray}{No}  \\ 
 PJ10  S & 2017-12-16  19:36:38 &48& 61\% & -900  $\pm$  300 &Yes&Yes& \textcolor{gray}{No} &Yes& \textcolor{gray}{No} &Yes&Yes&Yes \\ 
 PJ11  N & 2018-02-07  14:00:10 &259& 85\% & 300  $\pm$  300 &Yes&Yes& \textcolor{gray}{No} & \textcolor{gray}{No} & \textcolor{gray}{No} & \textcolor{gray}{No} & \textcolor{gray}{No} & \textcolor{gray}{No}  \\ 
 PJ11  S & 2018-02-07  15:43:53 &322& 82\% & 800  $\pm$  300 &Yes&Yes&Yes&Yes&Yes& \textcolor{gray}{No} &Yes& \textcolor{gray}{No}  \\ 
 PJ12  N & 2018-04-01  08:57:02 &131& 91\% & -1300  $\pm$  300 &Yes&Yes& \textcolor{gray}{No} & \textcolor{gray}{No} & \textcolor{gray}{No} & \textcolor{gray}{No} & \textcolor{gray}{No} & \textcolor{gray}{No}  \\ 
 PJ12  S & 2018-04-01  11:25:10 &220& 88\% & -1000  $\pm$  300 &Yes&Yes&Yes&Yes&Yes&Yes&Yes&Yes \\ 
 PJ13  N & 2018-05-24  05:22:26 &55& 65\% & 600  $\pm$  300 &Yes&Yes& \textcolor{gray}{No} & \textcolor{gray}{No} & \textcolor{gray}{No} & \textcolor{gray}{No} & \textcolor{gray}{No} & \textcolor{gray}{No}  \\ 
 PJ13  S & 2018-05-24  07:19:34 &126& 75\% & 900  $\pm$  300 &Yes&Yes& \textcolor{gray}{No} &Yes& \textcolor{gray}{No} & \textcolor{gray}{No} &Yes& \textcolor{gray}{No}  \\ 
 PJ14  N & 2018-07-16  05:03:39 &98& 79\% & 200  $\pm$  300 &Yes&Yes& \textcolor{gray}{No} & \textcolor{gray}{No} & \textcolor{gray}{No} & \textcolor{gray}{No} & \textcolor{gray}{No} & \textcolor{gray}{No}  \\ 
 PJ14  S & 2018-07-16  06:57:13 &167& 86\% & 500  $\pm$  300 &Yes&Yes&Yes&Yes&Yes&Yes&Yes&Yes \\ 
 PJ15  N & 2018-09-07  01:12:54 &13& 51\% & -1200  $\pm$  300 &Yes&Yes& \textcolor{gray}{No} & \textcolor{gray}{No} & \textcolor{gray}{No} & \textcolor{gray}{No} & \textcolor{gray}{No} & \textcolor{gray}{No}  \\ 
 PJ15  S & 2018-09-07  02:50:28 &72& 52\% & -500  $\pm$  300 &Yes&Yes& \textcolor{gray}{No} &Yes&Yes& \textcolor{gray}{No} &Yes& \textcolor{gray}{No}  \\ 
 PJ16  N & 2018-10-29  21:09:47 &280& 78\% & 400  $\pm$  300 &Yes&Yes& \textcolor{gray}{No} & \textcolor{gray}{No} & \textcolor{gray}{No} & \textcolor{gray}{No} & \textcolor{gray}{No} & \textcolor{gray}{No}  \\ 
 PJ16  S & 2018-10-29  22:45:00 &338& 63\% & 400  $\pm$  300 &Yes&Yes& \textcolor{gray}{No} &Yes&Yes&Yes&Yes&Yes \\ 
 PJ17  N & 2018-12-21  17:02:10 &185& 64\% & -400  $\pm$  300 &Yes&Yes& \textcolor{gray}{No} & \textcolor{gray}{No} & \textcolor{gray}{No} & \textcolor{gray}{No} & \textcolor{gray}{No} & \textcolor{gray}{No}  \\ 
 PJ17  S & 2018-12-21  18:54:49 &253& 77\% & 200  $\pm$  300 &Yes&Yes& \textcolor{gray}{No} & \textcolor{gray}{No} &Yes&Yes&Yes&Yes \\ 
 PJ18  N & 2019-02-12  17:43:36 &264& 0\% &-& \textcolor{gray}{No} & \textcolor{gray}{No} & \textcolor{gray}{No} & \textcolor{gray}{No} & \textcolor{gray}{No} & \textcolor{gray}{No} & \textcolor{gray}{No} & \textcolor{gray}{No}  \\ 
 PJ18  S & 2019-02-12  19:14:17 &319& 66\% & -700  $\pm$  300 & \textcolor{gray}{No} &Yes& \textcolor{gray}{No} &Yes& \textcolor{gray}{No} &Yes&Yes&Yes \\ 
 PJ19  N & 2019-04-06  11:55:26 &108& 41\% & 200  $\pm$  300 &Yes&Yes& \textcolor{gray}{No} & \textcolor{gray}{No} & \textcolor{gray}{No} & \textcolor{gray}{No} & \textcolor{gray}{No} & \textcolor{gray}{No}  \\ 
 PJ19  S & 2019-04-06  13:54:35 &180& 51\% & 700  $\pm$  200 &Yes&Yes& \textcolor{gray}{No} & \textcolor{gray}{No} & \textcolor{gray}{No} &Yes&Yes&Yes \\ 
 PJ20  N & 2019-05-29  08:07:26 &24& 85\% & -200  $\pm$  300 &Yes&Yes& \textcolor{gray}{No} & \textcolor{gray}{No} & \textcolor{gray}{No} & \textcolor{gray}{No} & \textcolor{gray}{No} & \textcolor{gray}{No}  \\ 
 PJ20  S & 2019-05-29  09:48:51 &86& 99\% & 100  $\pm$  300 &Yes&Yes&Yes&Yes& \textcolor{gray}{No} & \textcolor{gray}{No} &Yes& \textcolor{gray}{No}  \\ 
 PJ21  N & 2019-07-21  04:06:09 &293& 78\% & -700  $\pm$  300 &Yes&Yes& \textcolor{gray}{No} & \textcolor{gray}{No} & \textcolor{gray}{No} & \textcolor{gray}{No} & \textcolor{gray}{No} & \textcolor{gray}{No}  \\ 
 PJ21  S & 2019-07-21  05:59:19 &1& 95\% & -400  $\pm$  300 &Yes&Yes&Yes&Yes& \textcolor{gray}{No} &Yes&Yes&Yes \\ 
 PJ22  N & 2019-09-12  03:45:30 &334& 95\% & 1800  $\pm$  300 &Yes&Yes& \textcolor{gray}{No} & \textcolor{gray}{No} & \textcolor{gray}{No} & \textcolor{gray}{No} & \textcolor{gray}{No} & \textcolor{gray}{No}  \\ 
 PJ22  S & 2019-09-12  05:20:05 &31& 76\% & 1800  $\pm$  300 &Yes&Yes& \textcolor{gray}{No} &Yes&Yes&Yes&Yes&Yes \\ 
 PJ23  N & 2019-11-03  22:34:24 &200& 49\% & 0  $\pm$  300 &Yes&Yes& \textcolor{gray}{No} & \textcolor{gray}{No} & \textcolor{gray}{No} & \textcolor{gray}{No} & \textcolor{gray}{No} & \textcolor{gray}{No}  \\ 
 PJ23  S & 2019-11-04  00:14:55 &261& 94\% & 100  $\pm$  300 &Yes&Yes&Yes&Yes& \textcolor{gray}{No} & \textcolor{gray}{No} &Yes& \textcolor{gray}{No}  \\ 
 PJ24  N & 2019-12-26  17:23:52 &67& 51\% & 200  $\pm$  300 &Yes&Yes& \textcolor{gray}{No} & \textcolor{gray}{No} & \textcolor{gray}{No} & \textcolor{gray}{No} & \textcolor{gray}{No} & \textcolor{gray}{No}  \\ 
 PJ24  S & 2019-12-26  20:08:37 &166& 84\% & 600  $\pm$  300 &Yes&Yes&Yes&Yes& \textcolor{gray}{No} &Yes&Yes&Yes \\ 
 PJ25  N & 2020-02-17  18:08:03 &147& 0\% &-& \textcolor{gray}{No} & \textcolor{gray}{No} & \textcolor{gray}{No} & \textcolor{gray}{No} & \textcolor{gray}{No} & \textcolor{gray}{No} & \textcolor{gray}{No} & \textcolor{gray}{No}  \\ 
 PJ25  S & 2020-02-17  19:52:29 &210& 44\% & 400  $\pm$  300 & \textcolor{gray}{No} &Yes& \textcolor{gray}{No} &Yes&Yes&Yes&Yes&Yes \\ 
 PJ26  N & 2020-04-10  13:42:22 &41& 75\% & -1600  $\pm$  300 &Yes&Yes& \textcolor{gray}{No} & \textcolor{gray}{No} & \textcolor{gray}{No} & \textcolor{gray}{No} & \textcolor{gray}{No} & \textcolor{gray}{No}  \\ 
 PJ26  S & 2020-04-10  15:27:32 &104& 57\% & -700  $\pm$  300 &Yes&Yes& \textcolor{gray}{No} &Yes&Yes& \textcolor{gray}{No} &Yes& \textcolor{gray}{No}  \\ 
 PJ27  N & 2020-06-02  10:22:50 &334& 38\% & -900  $\pm$  300 &Yes&Yes& \textcolor{gray}{No} & \textcolor{gray}{No} & \textcolor{gray}{No} & \textcolor{gray}{No} & \textcolor{gray}{No} & \textcolor{gray}{No}  \\ 
 PJ27  S & 2020-06-02  12:21:20 &46& 49\% & -200  $\pm$  300 &Yes&Yes& \textcolor{gray}{No} &Yes& \textcolor{gray}{No} &Yes&Yes&Yes \\ 
 PJ28  N & 2020-07-25  06:25:33 &245& 12\% & 2300  $\pm$  300 &Yes& \textcolor{gray}{No} & \textcolor{gray}{No} & \textcolor{gray}{No} & \textcolor{gray}{No} & \textcolor{gray}{No} & \textcolor{gray}{No} & \textcolor{gray}{No}  \\ 
 PJ28  S & 2020-07-25  07:54:55 &299& 66\% & 2100  $\pm$  300 &Yes&Yes& \textcolor{gray}{No} &Yes& \textcolor{gray}{No} & \textcolor{gray}{No} &Yes& \textcolor{gray}{No}  \\ 
 PJ29  N & 2020-09-16  02:25:22 &153& 4\% & -1500  $\pm$  200 &Yes& \textcolor{gray}{No} & \textcolor{gray}{No} & \textcolor{gray}{No} & \textcolor{gray}{No} & \textcolor{gray}{No} & \textcolor{gray}{No} & \textcolor{gray}{No}  \\ 
 PJ29  S & 2020-09-16  04:23:42 &225& 90\% & -300  $\pm$  300 &Yes&Yes&Yes&Yes& \textcolor{gray}{No} & \textcolor{gray}{No} &Yes& \textcolor{gray}{No}  \\ 
 PJ30  N & 2020-11-08  01:57:01 &190& 9\% & 3000  $\pm$  300 &Yes&Yes& \textcolor{gray}{No} & \textcolor{gray}{No} & \textcolor{gray}{No} & \textcolor{gray}{No} & \textcolor{gray}{No} & \textcolor{gray}{No}  \\ 
 PJ30  S & 2020-11-08  03:28:47 &246& 76\% & 1000  $\pm$  300 &Yes&Yes& \textcolor{gray}{No} &Yes& \textcolor{gray}{No} &Yes&Yes&Yes \\ 
 PJ31  N & 2020-12-30  21:59:18 &100& 0\% &-& \textcolor{gray}{No} & \textcolor{gray}{No} & \textcolor{gray}{No} & \textcolor{gray}{No} & \textcolor{gray}{No} & \textcolor{gray}{No} & \textcolor{gray}{No} & \textcolor{gray}{No}  \\ 
 PJ31  S & 2020-12-30  23:43:37 &163& 82\% & 100  $\pm$  300 & \textcolor{gray}{No} &Yes&Yes& \textcolor{gray}{No} & \textcolor{gray}{No} & \textcolor{gray}{No} &Yes& \textcolor{gray}{No}  \\ 
 PJ32  N & 2021-02-21  17:36:05 &355& 71\% & -900  $\pm$  300 &Yes&Yes& \textcolor{gray}{No} & \textcolor{gray}{No} & \textcolor{gray}{No} & \textcolor{gray}{No} & \textcolor{gray}{No} & \textcolor{gray}{No}  \\ 
 PJ32  S & 2021-02-21  20:04:05 &84& 87\% & -1000  $\pm$  300 &Yes&Yes&Yes&Yes& \textcolor{gray}{No} &Yes&Yes&Yes \\ 
 PJ33  N & 2021-04-15  23:37:10 &267& 60\% & 0  $\pm$  300 &Yes&Yes& \textcolor{gray}{No} & \textcolor{gray}{No} & \textcolor{gray}{No} & \textcolor{gray}{No} & \textcolor{gray}{No} & \textcolor{gray}{No}  \\ 
 PJ33  S & 2021-04-16  01:55:31 &351& 77\% & -500  $\pm$  300 &Yes&Yes& \textcolor{gray}{No} &Yes& \textcolor{gray}{No} & \textcolor{gray}{No} &Yes& \textcolor{gray}{No}  \\ 
 PJ34  N & 2021-06-08  07:53:29 &261& 75\% & 800  $\pm$  300 &Yes&Yes& \textcolor{gray}{No} & \textcolor{gray}{No} & \textcolor{gray}{No} & \textcolor{gray}{No} & \textcolor{gray}{No} & \textcolor{gray}{No}  \\ 
 PJ34  S & 2021-06-08  09:45:58 &329& 80\% & 300  $\pm$  300 &Yes&Yes&Yes&Yes& \textcolor{gray}{No} & \textcolor{gray}{No} &Yes& \textcolor{gray}{No}  \\ 
 PJ35  N & 2021-07-21  08:14:03 &262& 64\% & 600  $\pm$  300 &Yes&Yes& \textcolor{gray}{No} & \textcolor{gray}{No} & \textcolor{gray}{No} & \textcolor{gray}{No} & \textcolor{gray}{No} & \textcolor{gray}{No}  \\ 
 PJ35  S & 2021-07-21  10:15:36 &336& 73\% & 100  $\pm$  300 &Yes&Yes& \textcolor{gray}{No} &Yes& \textcolor{gray}{No} &Yes&Yes&Yes \\ 
 PJ36  N & 2021-09-02  22:40:39 &55& 48\% & 300  $\pm$  300 &Yes&Yes& \textcolor{gray}{No} & \textcolor{gray}{No} & \textcolor{gray}{No} & \textcolor{gray}{No} & \textcolor{gray}{No} & \textcolor{gray}{No}  \\ 
 PJ36  S & 2021-09-03  00:43:55 &130& 98\% & 900  $\pm$  300 &Yes&Yes&Yes&Yes& \textcolor{gray}{No} &Yes&Yes&Yes \\ 
 PJ37  N & 2021-10-16  17:12:14 &357& 70\% & -200  $\pm$  300 &Yes&Yes& \textcolor{gray}{No} & \textcolor{gray}{No} & \textcolor{gray}{No} & \textcolor{gray}{No} & \textcolor{gray}{No} & \textcolor{gray}{No}  \\ 
 PJ37  S & 2021-10-16  18:53:35 &58& 98\% & 0  $\pm$  300 &Yes&Yes&Yes&Yes& \textcolor{gray}{No} &Yes&Yes&Yes \\ 
 PJ38  N & 2021-11-29  14:12:04 &27& 86\% & -200  $\pm$  300 &Yes&Yes& \textcolor{gray}{No} & \textcolor{gray}{No} & \textcolor{gray}{No} & \textcolor{gray}{No} & \textcolor{gray}{No} & \textcolor{gray}{No}  \\ 
 PJ38  S & 2021-11-29  16:13:59 &101& 91\% & 300  $\pm$  300 &Yes&Yes&Yes&Yes&Yes&Yes&Yes&Yes \\ 
 PJ39  N & 2022-01-12  10:33:08 &35& 16\% & -1100  $\pm$  300 &Yes& \textcolor{gray}{No} & \textcolor{gray}{No} & \textcolor{gray}{No} & \textcolor{gray}{No} & \textcolor{gray}{No} & \textcolor{gray}{No} & \textcolor{gray}{No}  \\ 
 PJ39  S & 2022-01-12  12:55:09 &120& 64\% & 500  $\pm$  300 &Yes&Yes& \textcolor{gray}{No} & \textcolor{gray}{No} & \textcolor{gray}{No} &Yes&Yes&Yes \\ 
 PJ40  N & 2022-02-25  02:01:00 &225& 32\% & -1500  $\pm$  300 &Yes&Yes& \textcolor{gray}{No} & \textcolor{gray}{No} & \textcolor{gray}{No} & \textcolor{gray}{No} & \textcolor{gray}{No} & \textcolor{gray}{No}  \\ 
 PJ40  S & 2022-02-25  03:55:23 &294& 61\% & -400  $\pm$  300 &Yes&Yes& \textcolor{gray}{No} &Yes&Yes& \textcolor{gray}{No} &Yes& \textcolor{gray}{No}  \\ 
 PJ41  N & 2022-04-09  15:50:02 &355& 76\% & 200  $\pm$  300 &Yes&Yes& \textcolor{gray}{No} & \textcolor{gray}{No} & \textcolor{gray}{No} & \textcolor{gray}{No} & \textcolor{gray}{No} & \textcolor{gray}{No}  \\ 
 PJ41  S & 2022-04-09  18:13:06 &81& 91\% & 400  $\pm$  300 &Yes&Yes&Yes&Yes&Yes& \textcolor{gray}{No} &Yes& \textcolor{gray}{No}  \\ 
 PJ42  N & 2022-05-23  02:17:38 &3& 36\% & 700  $\pm$  300 &Yes&Yes& \textcolor{gray}{No} & \textcolor{gray}{No} & \textcolor{gray}{No} & \textcolor{gray}{No} & \textcolor{gray}{No} & \textcolor{gray}{No}  \\ 
 PJ42  S & 2022-05-23  04:15:13 &75& 69\% & -300  $\pm$  300 &Yes&Yes& \textcolor{gray}{No} &Yes& \textcolor{gray}{No} &Yes&Yes&Yes \\ 
 PJ43  N & 2022-07-05  08:57:25 &234& 0\% &-& \textcolor{gray}{No} & \textcolor{gray}{No} & \textcolor{gray}{No} & \textcolor{gray}{No} & \textcolor{gray}{No} & \textcolor{gray}{No} & \textcolor{gray}{No} & \textcolor{gray}{No}  \\ 
 PJ43  S & 2022-07-05  13:45:48 &49& 32\% & -1600  $\pm$  300 & \textcolor{gray}{No} & \textcolor{gray}{No} & \textcolor{gray}{No} & \textcolor{gray}{No} & \textcolor{gray}{No} & \textcolor{gray}{No} &Yes& \textcolor{gray}{No}  \\ 
 PJ44  N & 2022-08-17  15:08:57 &88& 0\% &-& \textcolor{gray}{No} & \textcolor{gray}{No} & \textcolor{gray}{No} & \textcolor{gray}{No} & \textcolor{gray}{No} & \textcolor{gray}{No} & \textcolor{gray}{No} & \textcolor{gray}{No}  \\ 
 PJ44  S & 2022-08-17  17:19:53 &167& 98\% & 100  $\pm$  300 & \textcolor{gray}{No} &Yes&Yes&Yes& \textcolor{gray}{No} & \textcolor{gray}{No} &Yes& \textcolor{gray}{No}  \\ 
 PJ45  N & 2022-09-29  17:35:46 &166& 0\% &-& \textcolor{gray}{No} & \textcolor{gray}{No} & \textcolor{gray}{No} & \textcolor{gray}{No} & \textcolor{gray}{No} & \textcolor{gray}{No} & \textcolor{gray}{No} & \textcolor{gray}{No}  \\ 
 PJ45  S & 2022-09-29  19:33:52 &237& 96\% & 300  $\pm$  300 & \textcolor{gray}{No} &Yes&Yes&Yes& \textcolor{gray}{No} & \textcolor{gray}{No} &Yes& \textcolor{gray}{No}  \\ 
 PJ46  N & 2022-11-06  21:21:19 &259& 13\% & 1300  $\pm$  300 &Yes& \textcolor{gray}{No} & \textcolor{gray}{No} & \textcolor{gray}{No} & \textcolor{gray}{No} & \textcolor{gray}{No} & \textcolor{gray}{No} & \textcolor{gray}{No}  \\ 
 PJ46  S & 2022-11-06  23:38:56 &342& 54\% & 1000  $\pm$  300 &Yes&Yes& \textcolor{gray}{No} &Yes& \textcolor{gray}{No} &Yes&Yes&Yes \\ 
 PJ47  N & 2022-12-15  03:40:45 &85& 15\% & 800  $\pm$  300 &Yes& \textcolor{gray}{No} & \textcolor{gray}{No} & \textcolor{gray}{No} & \textcolor{gray}{No} & \textcolor{gray}{No} & \textcolor{gray}{No} & \textcolor{gray}{No}  \\ 
 PJ47  S & 2022-12-15  05:12:33 &141& 81\% & 1000  $\pm$  300 &Yes&Yes&Yes& \textcolor{gray}{No} & \textcolor{gray}{No} &Yes&Yes&Yes \\ 
 PJ48  N & 2023-01-22  05:54:42 &123& 0\% &-& \textcolor{gray}{No} & \textcolor{gray}{No} & \textcolor{gray}{No} & \textcolor{gray}{No} & \textcolor{gray}{No} & \textcolor{gray}{No} & \textcolor{gray}{No} & \textcolor{gray}{No}  \\ 
 PJ48  S & 2023-01-22  08:08:03 &204& 84\% & -600  $\pm$  300 & \textcolor{gray}{No} &Yes&Yes&Yes&Yes& \textcolor{gray}{No} &Yes& \textcolor{gray}{No}  \\ 
 PJ49  N & 2023-03-01  06:09:16 &89& 0\% &-& \textcolor{gray}{No} & \textcolor{gray}{No} & \textcolor{gray}{No} & \textcolor{gray}{No} & \textcolor{gray}{No} & \textcolor{gray}{No} & \textcolor{gray}{No} & \textcolor{gray}{No}  \\ 
 PJ49  S & 2023-03-01  08:32:36 &175& 67\% & 200  $\pm$  300 & \textcolor{gray}{No} &Yes& \textcolor{gray}{No} & \textcolor{gray}{No} & \textcolor{gray}{No} &Yes&Yes&Yes \\ 
 PJ50  N & 2023-04-08  08:36:41 &135& 0\% &-& \textcolor{gray}{No} & \textcolor{gray}{No} & \textcolor{gray}{No} & \textcolor{gray}{No} & \textcolor{gray}{No} & \textcolor{gray}{No} & \textcolor{gray}{No} & \textcolor{gray}{No}  \\ 
 PJ50  S & 2023-04-08  10:35:05 &206& 83\% & -200  $\pm$  300 & \textcolor{gray}{No} &Yes&Yes&Yes& \textcolor{gray}{No} &Yes&Yes&Yes \\ 
 PJ51  N & 2023-05-16  07:38:52 &57& 0\% &-& \textcolor{gray}{No} & \textcolor{gray}{No} & \textcolor{gray}{No} & \textcolor{gray}{No} & \textcolor{gray}{No} & \textcolor{gray}{No} & \textcolor{gray}{No} & \textcolor{gray}{No}  \\ 
 PJ51  S & 2023-05-16  09:42:47 &132& 76\% & 500  $\pm$  300 & \textcolor{gray}{No} &Yes& \textcolor{gray}{No} & \textcolor{gray}{No} & \textcolor{gray}{No} & \textcolor{gray}{No} &Yes& \textcolor{gray}{No}  \\ 
 PJ53  N & 2023-07-31  09:29:34 &37& 0\% &-& \textcolor{gray}{No} & \textcolor{gray}{No} & \textcolor{gray}{No} & \textcolor{gray}{No} & \textcolor{gray}{No} & \textcolor{gray}{No} & \textcolor{gray}{No} & \textcolor{gray}{No}  \\ 
 PJ53  S & 2023-07-31  11:50:11 &122& 43\% & -100  $\pm$  300 & \textcolor{gray}{No} &Yes& \textcolor{gray}{No} & \textcolor{gray}{No} & \textcolor{gray}{No} & \textcolor{gray}{No} &Yes& \textcolor{gray}{No}  \\ 
 PJ54  N & 2023-09-07  12:05:54 &89& 0\% &-& \textcolor{gray}{No} & \textcolor{gray}{No} & \textcolor{gray}{No} & \textcolor{gray}{No} & \textcolor{gray}{No} & \textcolor{gray}{No} & \textcolor{gray}{No} & \textcolor{gray}{No}  \\ 
 PJ54  S & 2023-09-07  14:39:53 &182& 96\% & -1300  $\pm$  300 & \textcolor{gray}{No} &Yes&Yes& \textcolor{gray}{No} &Yes& \textcolor{gray}{No} &Yes& \textcolor{gray}{No}  \\

    \end{longtable}   
\end{ThreePartTable}

\end{landscape}

\centering

\begin{ThreePartTable}
    \centering
    \setlength\LTleft{0pt}
    \setlength\LTright{0pt}
    \setlength\tabcolsep{4pt}


    \begin{TableNotes}
      \item\footnotesize ``ID'' gives the unique identifier of the HST exposure. ``Date'' gives the average date and time of the HST exposure, in the format YYYY-MM-DD hh:mm:ss. ``$\phi_{SS}$'' denotes the average left-handed System-III subsolar longitude of Jupiter during the exposure. ``$\phi_{CML}$'' denotes the average left-handed System-III central meridian longitude (sub-terrestrial longitude) of Jupiter during the exposure. ``Expansion'' refers to the day-side (visible by HST) expansion of the ME relative to the reference contour. ME expansions are given as positive where the detected ME is, on average, equatorward of the reference contour. Negative values indicate global contraction of the ME compared to the reference contour. ``MS compressed?'' denotes whether the magnetosphere was in a state of compression (``Y'') or uncompressed (``N''), as determined by \citet{yao+:2022}. 
    \end{TableNotes}
   
    \begin{longtable}[htbp]{ llllll }
        \label{tab:hst_cases}\\
        \caption{HST-STIS cases used in this work.}\\
    
        \toprule
        ID & Date & 	$\phi_{SS}$ (\textdegree) & $\phi_{CML}$ (\textdegree) & Expansion (km) & MS compressed? \\
        \midrule
        \endhead

        \midrule[\heavyrulewidth]
        \multicolumn{6}{r}{\textit{continued...}}\\
        \endfoot

        \midrule[\heavyrulewidth]
        \insertTableNotes  
        \endlastfoot
   
    ocx837olq& 2016-06-26  01:39:14 &141&151& -800  $\pm$  700 &N \\ 
ocx838xzq& 2016-06-26  22:19:19 &170&181& -900  $\pm$  900 &N \\ 
ocx840t4q& 2016-06-30  04:13:09 &116&125& -3100  $\pm$  900 &Y \\ 
ocx844t7q& 2016-07-14  16:23:27 &142&151& -2000  $\pm$  900 &Y \\ 
ocx846gwq& 2016-07-17  14:21:11 &160&168& -1600  $\pm$  700 &Y \\ 
ocx842ewq& 2016-07-18  18:58:22 &118&126& -1400  $\pm$  700 &Y \\ 
ocx847eyq& 2016-07-18  20:33:46 &175&184& -2100  $\pm$  700 &Y \\ 
od8k25jlq& 2017-01-22  15:31:08 &203&193& 0  $\pm$  800 &N \\ 
od8k29fiq& 2017-01-23  20:07:50 &161&151& -600  $\pm$  800 &N \\ 
od8k30i4q& 2017-01-24  15:11:42 &132&122& -2000  $\pm$  800 &Y \\ 
od8k31iaq& 2017-01-24  16:47:04 &190&180& -1000  $\pm$  800 &Y \\ 
od8k55tnq& 2017-03-01  15:56:39 &179&172& -600  $\pm$  800 &N \\ 
od8k56o2q& 2017-03-05  18:29:08 &153&147& -500  $\pm$  800 &N \\ 
od8k32anq& 2017-03-17  08:39:23 &162&158& -1200  $\pm$  700 &Y \\ 
od8k42fiq& 2017-03-18  14:52:23 &178&174& -500  $\pm$  700 &Y \\ 
od8k57itq& 2017-03-19  09:57:00 &150&146& -1600  $\pm$  800 &Y \\ 
od8k82ovq& 2017-04-23  14:00:08 &163&167& -1200  $\pm$  800 &N \\ 
od8k41ixq& 2017-05-09  06:42:14 &146&152& -300  $\pm$  700 &N \\

    \end{longtable}   
\end{ThreePartTable}

\end{appendix}

\end{document}